\documentclass[pre,twocolumn,reprint]{revtex4}
\usepackage{amsmath,amssymb}

\usepackage{overpic,color}

\newcommand{\bu}{{\bf u}}
\newcommand{\br}{{\bf r}}

\newcommand{\intn}{\!\int\!}
\newcommand{\dd}{\mathrm{d}}

\usepackage{amssymb}
\usepackage{amsmath}
\usepackage[T1]{fontenc}
\usepackage{amsfonts}
\usepackage{graphicx}
\usepackage{subfigure}
\usepackage{dcolumn}
\usepackage{bm}
\usepackage{hyperref}
\usepackage{empheq}
\usepackage{natbib}
\usepackage{color}
\newcommand*{\citen}{}
\DeclareRobustCommand*{\citen}[1]{%
  \begingroup
    \romannumeral-`\x 
    \setcitestyle{numbers}%
    \cite{#1}%
  \endgroup
}

\begin{document}

\title{Defect patterns of two-dimensional nematic liquid crystals in confinement
}

\author{Xiaomei Yao and  Lei Zhang}
\affiliation {Beijing International Center for Mathematical Research, Peking University, Beijing
100871, China.}
\date{\today}

\author{Jeff Z. Y. Chen\footnote{Email: jeffchen@uwaterloo.ca}}
\affiliation{ Department of Physics and Astronomy, University of Waterloo, Waterloo, Ontario, Canada, N2L 3G1 }
\date{\today}

\begin{abstract} 
A two-dimensional or quasi-two-dimensional nematic liquid crystal
refers to a surface confined system.
When such a system is further confined by external line boundaries
or excluded from internal line boundaries, the nematic directors form a deformed texture that may display
defect points or defect lines, for which winding numbers can be clearly defined.
Here, a particular attention is paid to the case
 when the liquid crystal molecules prefer to form a boundary nematic texture in parallel to the wall surface (i.e., following the homogeneous boundary condition).
A general theory, based on geometric argument,  is presented for the
relationship between the sum of all winding numbers in the system (the total winding number)  and the type of confinement angles and curved segments.
The conclusion is validated by comparing the theoretical defect rule with existing nematic textures observed experimentally and theoretically in recent years.
\end{abstract}

\maketitle

\clearpage
\newpage
\section{\bf INTRODUCTION}

The bulk state of a nematic liquid crystal is
a spatially uniform fluid with the long molecular axes spontaneously
ordered in a common direction, for which a global nematic director can be defined. A familiar example
is a nematic system composed of rodlike linear molecules, for which
the bulk state can be described by a uniform nematic field in one direction. In an ideal nematic state, the field lines, similar
to the field lines of a uniform electric field, extend in space. For a nonpolar system, the case considered here, these nematic field
lines have no arrows (i.e., are head-to-tail symmetry) \cite{deG1993}.

The introduction of a physical boundary, however, disrupts the otherwise uniform nematic field lines. Depending on how molecules are
aligned at the boundary, this could create a frustration on the nematic field lines. When liquid crystal molecules prefer to align along  the wall surface, a so-call ``homogeneous'' liquid crystal boundary surface is formed. The nematic field lines would then line up with the surface conditions and in the mean time, attempt to keep minimal field-line deformation to lower the distortion elastic energies. In a finite system, according to the physical conditions, the nematic field lines can form defect points, lines, etc.

For example, a well-studied problem is the understanding of the structures formed by colloid particles (mostly of the spherical shape) emersed in a liquid crystal. Tremendous experimental and theoretical progress has been made in the last two decades, devoting to such systems \cite{Poulin1997, Nazarenko2001, Muvsevivc2006two, vskarabot2007two, ognysta20082d, Ravnik2007entangled, vskarabot2008interactions, Ravnik2009braids, Ognysta2011, Tkalec2013topology,wang2017,Wang2018}. The presence of colloid particle surfaces induces liquid crystal defects and in turn, the tendency of minimizing the overall liquid-crystal defect free-energy couples the colloid particles in a particular form, yielding, e.g., ordered three-dimensional colloidal crystals.

Another commonly studied example is a liquid crystal fluid confined to curved surfaces, which has drawn significant theoretical and experimental
attentions in recent years. Depending on the geometry of the confining surface, the system may display both density and orientational field defects, which can be detected experimentally\cite{Lubensky1992,Nelson2002a,arsenault2004towards,li2009site,Bowick2009,fernandez2007noval,Lopez-Leon2011}.
The nature of an ordered state depends on the geometric parameters as wells as how far the system is away from the isotropic-nematic transition. Commonly used examples in theories and computer simulations are the nematic defect structures formed by a two-dimensional fluid containing liquid-crystal molecules confined on a spherical surface\citep{Lubensky1992,nelson2002toward,huber2005,skacej2008controlling, Shin2008,Bates2008,dhakal2002nematic,Zhang2012prl,Zhang2012,LiMiaoMaChen2013,Liang2014}, or a toroidal surface
\cite{Evans1995,Bowick2004,Selinger2011,LiYao2014,Segatti2014,Jesenek2015}.
A number of theoretical and computer-simulation approaches have been taken to study nematic structures in confinement. The Frank elastic and Landau-de Gennes free-energy models are often used and contain phenomenological parameters \cite{Sheng1976,Sheng1982,Lubensky1992,nelson2002toward,huber2005,skacej2008controlling,dhakal2002nematic,Napoli2012,Napoli2012a,LiYao2014}.
The molecular-level based models, either the simpler Onsager and Maier-Saupe theories, or the more complicated density-functional theories, contain system parameters that can be traced back to the physical origins \cite{Chrzanowska2001,Chrzanowska2003,delasHeras2004,delasHeras2009,Emelyanenko2011,Zhang2012prl,Zhang2012,Chen2013,Liang2014}.
A universal mathematical theorem is such that  the Euler characteristics of the confining manifold uniquely determine the total
 winding numbers associated with the liquid crystal defects. For example,  
 the embedded director fields have either total winding number 2, or 0, for a spherical or colloidal surface, respectively.
  
Less, however, is understood about the general feature of another type of confinement. 
The past two decades have witnessed a surge in research activities on the topic of
boundary-frustrated nematic states, when a liquid-crystal-like system in two dimensions (2D) is confined by a
closed boundary line. These systems can be a flat square box containing traditional
liquid crystal molecules \cite{Tsakonas2007}, a collection of
visible steel needles in circular and square boxes which are equilibrated by a vibrational bed
\cite{Galanis2006Spontaneous}, micron-sized rodlike colloid particles in confinement \cite{Cortes2017,Aarts2021Particle},
or even semiflexible biological molecules confined in chambers of various shapes \cite{Mulder2011,Lewis2014,Garlea2019Colloidal}, all
under physical conditions that can be classified as in quasi 2D.
On the theoretical side, various theoretical approaches have been undertaken to model related systems, including solving
Oseen-Frank (OF) model \cite{Lewis2014,han2021}, Landau-de~Gennes (LdG) model \cite{Tsakonas2007,Everts2016A,Majumdar2017,Wang2018Order,Yin2020Construction,Han2020A,Han2021Solution}, and the density-functional theories such as the Onsager model \cite{Chen2013,Yao2018,Yao2020}
or beyond \cite{Aarts2021Particle}. Monte Carlo (MC)
simulations of confined rigid molecules in 2D or quasi 2D
have also been made \cite{Dzubiella2000Topological,Heras2014Domain,Mulder2015,Garlea2016Finite,Garlea2019Colloidal,Hashemi2019Structure,Hashemi2019Finding}.
A rich variety of defect patterns have been obtained from these studies.

Hence, we face a fundamental question: how to set up a universal theory that can be used to explain
 the total defect winding number found in these 2D, line-confined systems,  regardless of the experimental and theoretical methods used.
  The current paper serves two purposes. First, we derive the defect rules by using the characteristics of
  the confining boundaries. The simple case of a nematic fluid confined by an outer boundary is considered first in Sect.~\ref{W1},
  then followed by the case of a bulk nematic fluid containing an obstacle in Sect.~\ref{W2}]. The general defect rule for more complicated confining geometries, also covering the above two cases, is then generalized in Sect. \ref{W3}.

\begin{figure*}[!t]
\centering
  \includegraphics[width=2\columnwidth]{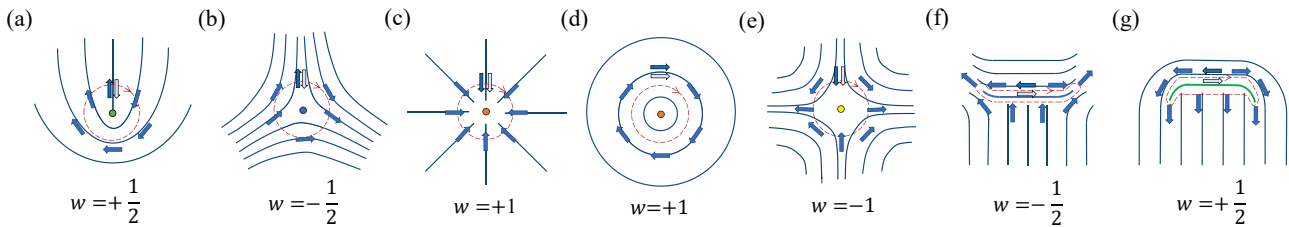}
  \caption{Schematics of local defect types with winding numbers (a) $w=+1/2$, (b) $w=-1/2$, (c, d) $w=+1$, (e) $w=-1$, (f) $w=-1/2$, and (g) $w=+1/2$,
   where (a) to (e) are patterns near defect points, and (f) as well as (g)
   patterns containing defect lines. The blue curves represent the directions of the (headless) nematic directors in space and the red dashed circles the path taken to evaluate the winding numbers. The same color scheme are used to illustrate the defect types in all the subsequent figures.
    \label{FIG1}}
\end{figure*}
\begin{figure}
\centering
  \includegraphics[width=0.95\columnwidth]{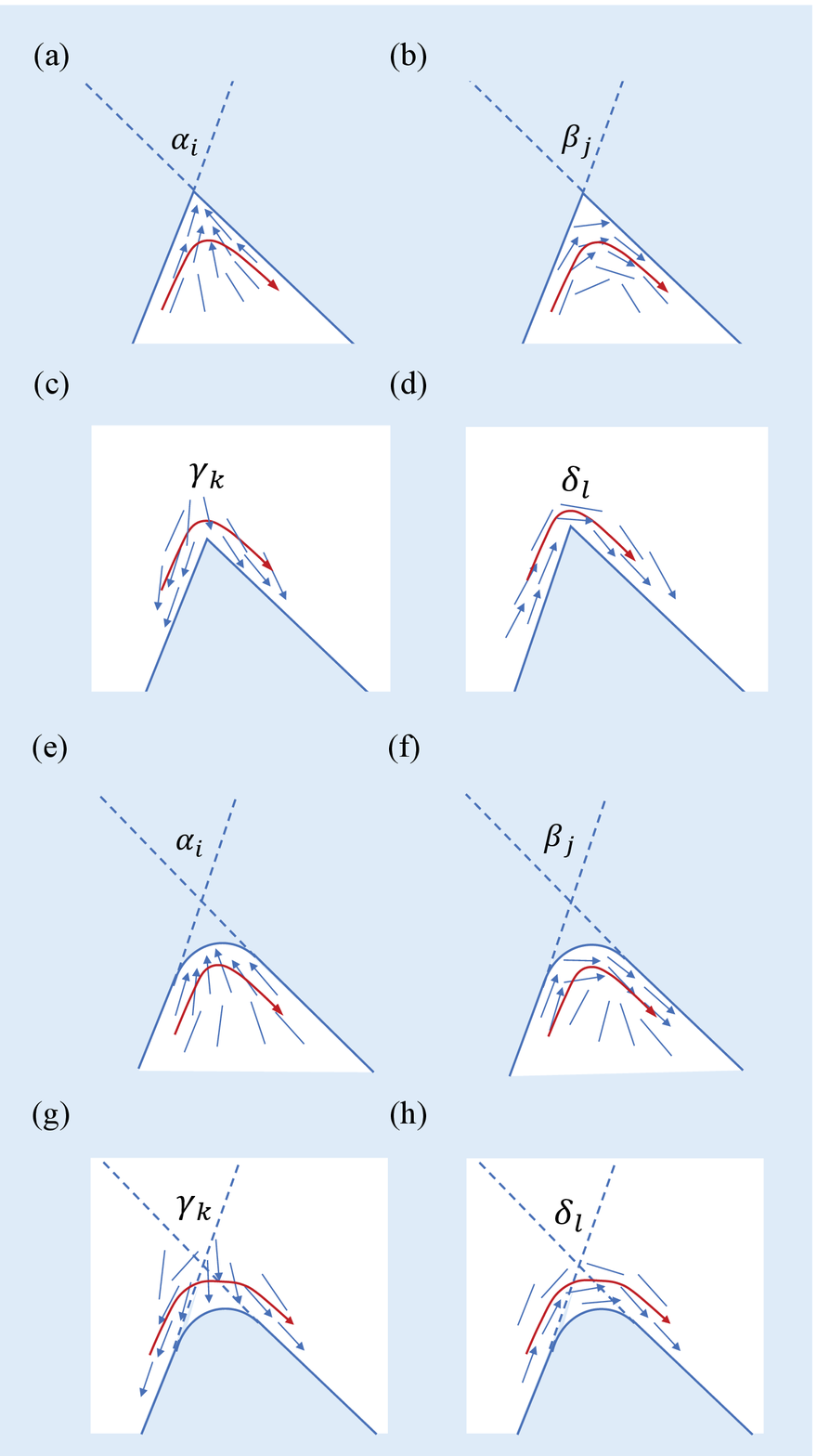}
  \caption{Basic types of nematic textures, containing splay and bend patterns.
   Shown in (a) and (b) are acute confinement corners, $\alpha_i$ and $\beta_j$,  that contain splay and bending textures; in (c) and (d) reflex corners, $\gamma_k$ and $\delta_l$,
  that contain splay and bend textures. Curved confinement segments in (e) to (h) can be dealt with by a  similar definition, by extending the linear segments in connection with a typical curve, to form an acute or reflex angle.}
  \label{FIG2}
\end{figure}

The second purpose is to comprehensively review the defect patterns discovered by various experimental, theoretical, and computer-simulation studies, in light of the defect rules that are deduced in this paper [see Sect. \ref{Comp}]. As summarized in
Table~\ref{Tab1}, most of the boundary conditions used in these studies have the geometrical shapes of  circles, triangles, rectangles, and hexagons. To supplement the existing studies, in Sects.~\ref{W1}, \ref{W2}, and \ref{W3}, we have provided
the defect patterns obtained from numerical
solutions to the Onsager model (see Appendix), for more complicated confinement types. A comparison between the defect patterns produced from
studies listed in Table~\ref{Tab1} and from our supplemented cases validates the general defect rules determined in the current work.

\section{TOTAL WINDING NUMBER INSIDE CONFINEMENT}\label{W1}

\subsection{Winding number of a single defect}\label{Wno}

For completeness, the definition of the winding number of a single defect is reviewed here. Figure~\ref{FIG1} illustrates some basic types of local defect patterns, where the blue curves connect local nematic directors. A complete spatial path is taken about the defect, shown by the clockwise, dashed red circle. Although the nematic directors are head-to-tail symmetry, blue arrows have been drawn for accounting purpose. As the red path completes its circle, the nematic director spins from the light blue arrow to the dark blue arrow; the
spinning angle, in units of $2\pi$, is defined as the winding number $w$. The sign of the winding number is positive if the arrow spins in the same direction as the red path, otherwise negative.

\subsection{Total winding number of defects in confinement: theory}

The summed, total winding number of all defects of a nematic liquid crystal confined
inside a polygon can be determined in a procedure similar to that in the last section. Instead of a local evaluation path around a defect point,
for this purpose, we take a complete path inside the boundary of a polygon and evaluate the spinning of the nematic director.

\begin{figure*}[!t]
\centering
  \includegraphics[width=2\columnwidth]{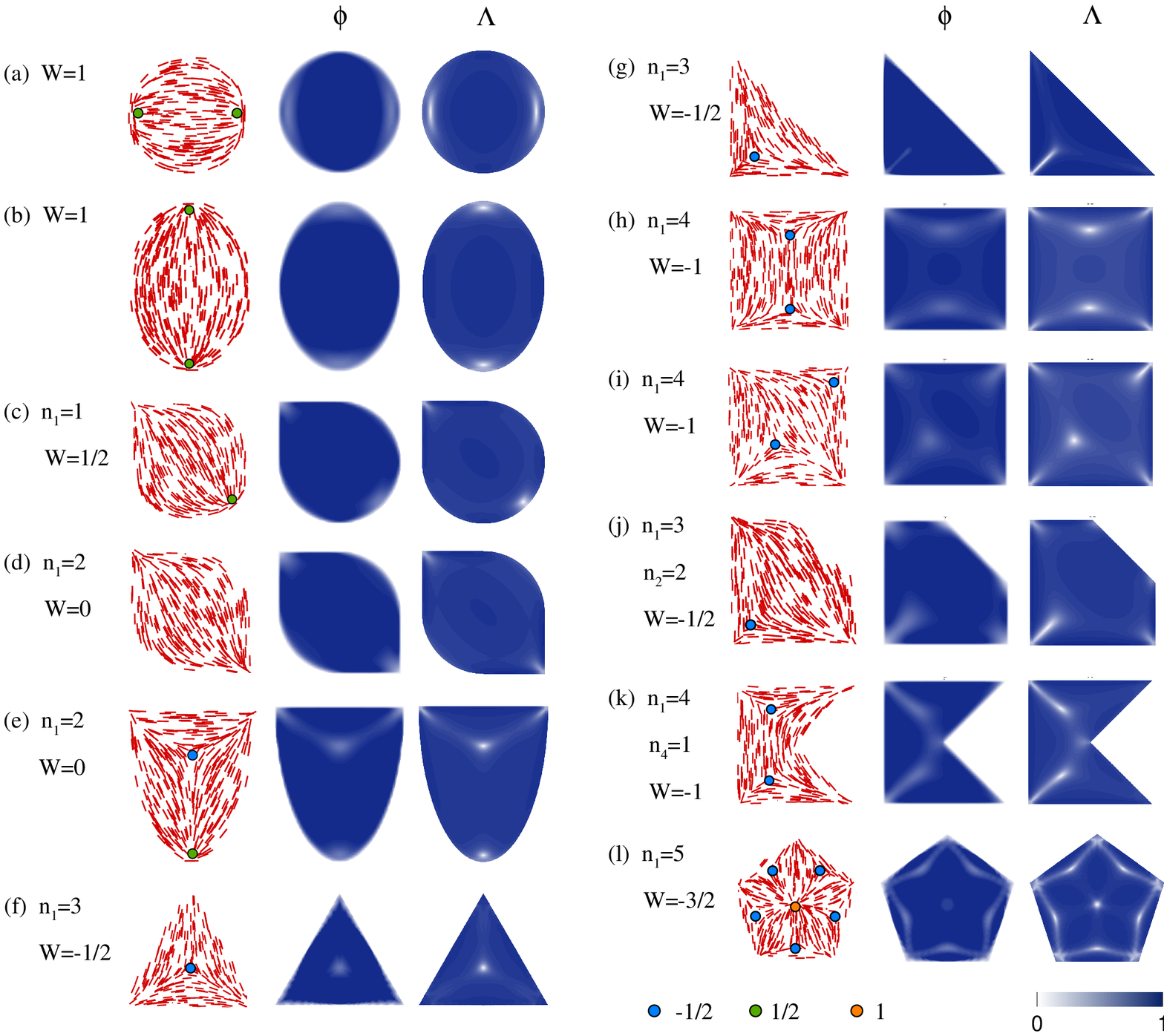}
  \caption{Examples of defect patterns obtained from the solution to the extended Onsager model.
  The reduced density used is $\tilde \rho = 10$
   and the reduced length $\tilde{L}$ in \eqref{Ldef} is in the range
  $[0.1, 0.25]$.
  The first column contains reconstructed schematic plots based on the numerical solution presented in other columns, where the defect points are indicated by colored circles.  The blue, green, and orange circles label the defect locations of -1/2, +1/2, and +1 winding numbers, respectively. The relative density $\phi(\mathbf r)$ and  orientational
   order parameter $\Lambda(\mathbf r)$ are plotted in columns 2, and 3, respectively. The values of $n_1, n_2, n_3$ and $n_4$ are omitted when they are equal to zero.
    \label{FIG3}}
\end{figure*}

The four basic types of nematic texture near polygonal corners are illustrated in Figs. \ref{FIG2} (a)-(d).
At a length scale much greater than the molecular dimension,
there are two typical nematic-director patterns, splay and bend, shown here in
Figs. \ref{FIG2}(a) and (b) inside an acute angle. As the evaluation red path passes around the corners, the nematic directors
spin by angles $-\alpha_i$ and $\pi-\beta_j$, in (a) and (b), respectively. The indices
 $i$ and $j$ have been added to denote the $i$th and $j$th acute angles that contain splay and bend textures, correspondingly.
In rare cases, the confinement geometry may contain a reflex angle, illustrated in Figs. \ref{FIG2}(c) and (d), for two typical patterns, splay and bend. As the evaluation red path takes place around a sharp boundary of an reflex angle, the nematic directors
of the $k$th splay and $l$th bend patterns spin by angles $2\pi - \gamma_k$ and $\pi - \delta_l$, respectively.

Around the interior of the confinement, assume that there are $n_1, n_2, n_3$ and $n_4$ angles of type (a), (b), (c) and (d) in
Fig. \ref{FIG2}, and other molecules near the boundary are parallel to the wall.
The complete evaluation path takes all these basic types and gives rise to a total winding number

\begin{equation}\label{Wdef}
\begin{aligned}
W &= {1\over 2\pi} \left [ \sum_{i=1}^{n_1} (-\alpha_i) + \sum_{j=1}^{n_2} (\pi-\beta_j)\right.\\
  &\left.+  \sum_{k=1}^{n_3} (2\pi-\gamma_k) +  \sum_{l=1}^{n_4} (\pi-\delta_l) \right ].
\end{aligned}
\end{equation}

According to geometric theory, the total sum of all angles inside a polygon of any shape is $(n_1+n_2+n_3+n_4-2)\pi$. This simplifies the above expression to
\begin{equation}\label{Winside}
W = -{1\over 2} (n_1- n_3-2).
\end{equation}
This is one of the main results of the current paper. Note that $W$ is determined by number of corners that contain
splay patterns only.

{One important generalization of the above expression is for nematic defects inside a closed boundary that
is either completely composed of or
partially contains
a curve. Taking a curve segment,
we can extend the tangent lines of the terminal ends of the curve to form a tangent angle, shown in Figs}. \ref{FIG2}(e)-(h). Depending on the types of nematic textures near the curved boundary, e.g. splay or bend, the above formula can be
directly used by counting number of splay patterns
associated with these curves.

\subsection{Examples}\label{sam}

In an earlier publication \cite{Yao2018}, examples of confinement boundaries formed by
acute angles, which were assumed to contain splay patterns only, and curve segments, which were assumed to
contain bend textures only, were examined. In such a case, $W$ in Eq.
\eqref{Winside} has a simpler version: $W = -{1\over 2} (n_1-2)$ where $n_1$ is the number of acute angles of the confinement boundary. Sect. \ref{SplayBend} further addresses the consistency of how a splay or bend angle is identified.

Here, we demonstrate the usefulness of the expression in \eqref{Winside} by examining the examples from the
numerical solutions to the extended
Onsager model for lyotropic nematic liquid crystals under various types of confinement. The model is
based on a classical free-energy model that Onsager developed for rodlike molecules of length $L$,
interacting with each other
through excluded volume interactions \cite{Onsager1949}. The onset of the bulk nematic state, in which no spatial variations
exist, depends on a single, reduced parameter
\begin{equation}\label{Rhodef}
\tilde \rho=\rho_0 L^2,
\end{equation}
where $\rho_0$ is the number of rodlike molecules per unit area.
When the model is extended to include spatial dependence and effects of the boundary conditions, it can be effectively used to model
a lyotropic liquid crystal in confinement, adding an additional system parameter
\begin{equation}\label{Ldef}
\tilde L = L/a,
\end{equation}
where $a$ is the typical size of the confinement geometry. More details can be found in Refs. \citen{Chen2013,Yao2018,Yao2020} and Appendix \ref{A1}.

A comparison between the defect rule in \eqref{Winside} and the numerical solutions can be viewed
in Fig. \ref{FIG3}, where the first column displays the reconstructed defect patterns according to the density profile $\rho({\mathbf r}, {\mathbf u})$, for direct visualization. The reduced density profile $\phi({\mathbf r})$, averaged over all orientational dependence $\mathbf u$ and normalized by $\rho_0$, is
a function of the spatial position specified by $\br$.
Displayed in the second column, depletion of the density can be clearly viewed around
the defect location. The orientational order parameter is assessed by
the order parameter tensor ${\mathsf Q}({\mathbf r})$ as a function of the spatial coordinates $\mathbf r$.
The average $\left< ...\right>$ is performed with respect to the angular dependence $\theta$ only.
In 2D, it is a $2\times 2$ traceless and symmetric tensor,
\begin{equation}\label{Qdef}
{\mathsf Q}({\mathbf r}) = \langle
{\mathbf u}{\mathbf u} - {\mathsf I}/2 \rangle
 = {1\over 2} \left[\begin{array}{cc}
      S({\mathbf r}) & T({\mathbf r})\\
      T({\mathbf r}) & -S({\mathbf r})
    \end{array}\right],\\
\end{equation}
where $\mathsf I$ is a unit tensor, and the right-hand side is the matrix representation of the tensor which contains the elements
$ S({\mathbf r}) = \left< \cos2\theta \right >$ and $ T({\mathbf r}) = \left< \sin2\theta \right >$, $\theta$ being the angle that
a rodlike molecule makes with respect to the horizontal axis. The eigenvalue of the $\mathsf Q$-tensor, $\Lambda = (S^2+T^2)^{1/2}$, is plotted in the third row, in which a defect point shows up at
a location where $\Lambda = 0$.

Among the plots, the circular and oval confinements in Figs. \ref{FIG3}(a) and (b) are two interesting cases. They can be viewed as smooth curves with zero angles, hence all $n_1=n_2=n_3=n_4=0$, which gives rise to $W=+1$ according to
\eqref{Winside}. In geometry, a circular shape could also be viewed as the asymptotic limit of a regular polygon when the polygon edge
number approaches infinity. From the latter perspective, Fig. \ref{FIG3}(a) corresponds to the case of $n_1=n_3=n_4=0$ but $n_2 \to \infty$; because the defect rule is not affected by the number of angles containing bend textures, $W=+1$.

The boundaries in (c)-(e) consist of lines and curves, and that in (f)-(l) contain polygonal segments. Most of the angles here are acute angles and liquid-crystal molecules prefer to align in splay patterns. In particular, there are two acute angles in (j) and a reflex angle in (k) around which nematic liquids are in bend patterns. In short summary, the defect rule deduced based on geometry consideration is fully consistent with the numerical solutions from an actual molecular theory.

\section{TOTAL WINDING NUMBER OF A NEMATIC LIQUID CONTAINING AN OBSTACLE}\label{W2}

\subsection{Total winding number of defects outside an obstacle: theory}

Here the case of a two-dimensional obstacle emersed in a nematic liquid is considered.
The basic defect
types in the liquid, near a corner angle
or a curved boundary, are the same as those illustrated in Fig. \ref{FIG2}.

Along the immediate exterior of the obstacle, a complete evaluation path
encounters $m_1$, $m_2$, $m_3$, and $m_4$ patterns of the type $\alpha_i$, $\beta_j$ $\gamma_k$
and $\delta_l$. The total winding number is hence the same as in \eqref{Wdef}, that is,
\begin{equation}\label{Wdef1}
\begin{aligned}
W &= {1\over 2\pi} \left [ \sum_{i=1}^{m_1} (-\alpha_i) + \sum_{j=1}^{m_2} (\pi-\beta_j)\right.\\
  &\left.+  \sum_{k=1}^{m_3} (2\pi-\gamma_k) +  \sum_{l=1}^{m_4} (\pi-\delta_l) \right ].
\end{aligned}
\end{equation}
The only difference is that the total sum of all angles outside a polygon of any shape is now $(m_1+m_2+m_3+m_4+2)\pi$, which makes
\begin{equation}\label{Woutside}
W =  -{1\over 2} (m_1- m_3 +2).
\end{equation}
In comparison with \eqref{Winside}, note the sign difference in front of $2$.
The total winding number is related to the number of angles (or the extended tangent angle from a curve segment) where
the nearby liquid displays splay nematic patterns.

\subsection{Examples}

\begin{figure}[!t]
\centering
  \includegraphics[width=1\columnwidth]{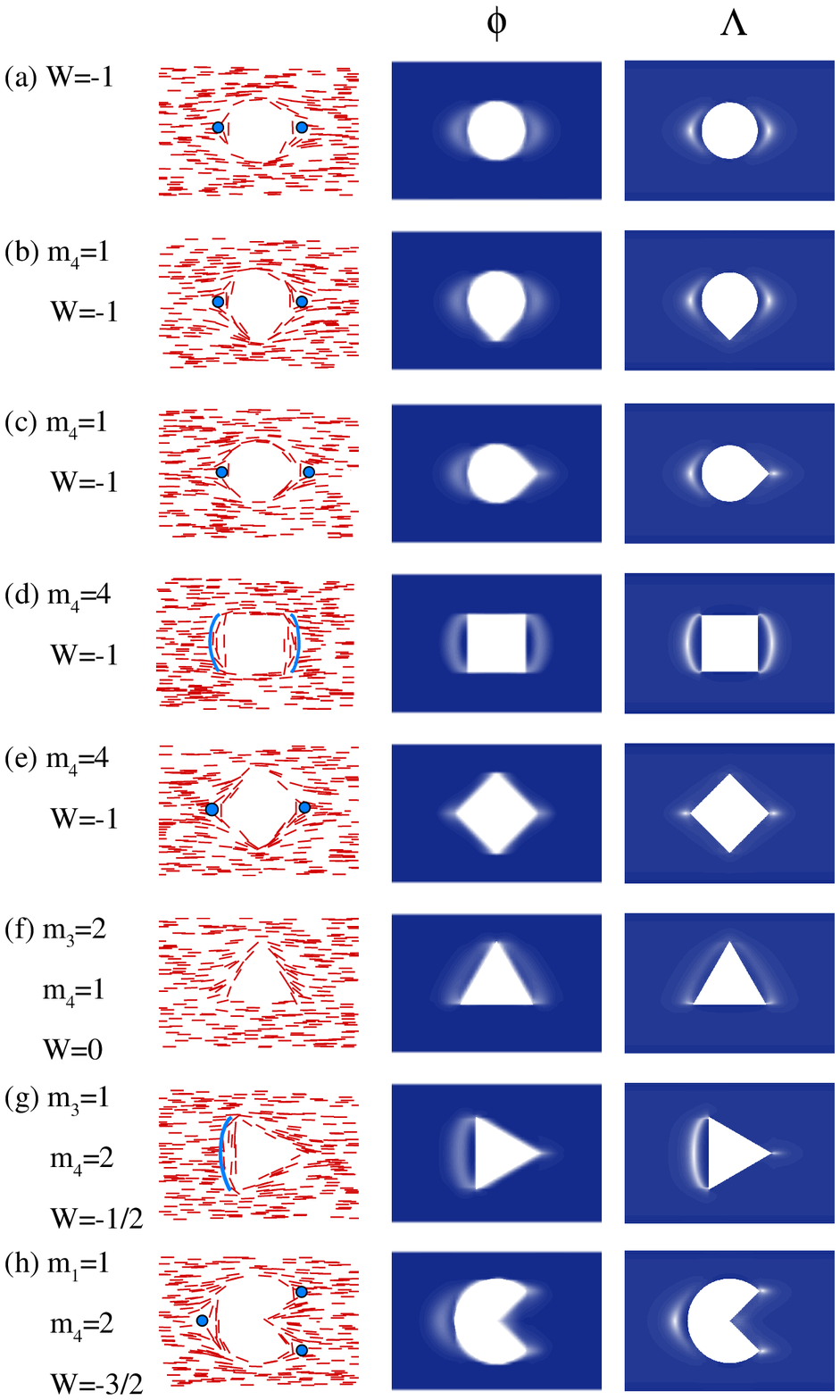}
  \caption{Examples of defect patterns obtained from the solution to the extended Onsager model.
The reduced density used is $\tilde \rho = 10$
and the reduced length $\tilde{L}=0.06$, defined in \eqref{Ldef}.
Colors used to specify the defect points and lines are the same as those in Figs. \ref{FIG2} and \ref{FIG3}. The non-zero values of $m_1, m_2, m_3$ and $m_4$, together with the total $W$, are specified in each plot.
    \label{FIG4}}
\end{figure}

Solving the same extended Onsager model, we obtain the numerical solutions for the density profiles
when differently shaped obstacles are immersed in an originally uniform nematic field.
A few produced examples are displayed in
Fig. \ref{FIG4}, in which the values of $m_1, m_2, m_3$ and $m_4$ used in the above defect rule are also shown.

The circular obstacle in (a) has no angles, hence all $m$'s vanish to yield $W=-1$.
Though
 the obstacles in (b) and (c) contain a sharp angle, the nearby nematic
 liquids make a bend pattern, hence $m_4=1$. The value of $m_4$, however, does not contribute to $W$ in \eqref{Woutside};
 this places them at the same category as (a) where $W=-1$.

 The square obstacle in (d) and (e) has four corners but the nematic liquid around them has a bending texture. Hence
 $m_4=4$, which makes $W=-1$. The nematic
 defects   in (d), though, are line defects, which can be contrasted with the defect points in (e).

 Plots (f) and (g) demonstrate that the geometric shape alone is not the determinant  factor that determines the value of $W$.
 Both obstacles are triangles but are placed in the nematic liquid in different orientations.
 In (f), the nematic liquid around the two lower corners displays splay patterns, hence $m_3=2$; the upper corner is associated with a bend texture, which gives $m_4=1$. In total, according to \eqref{Woutside}, $W=0$, which implies no defects in the nematic liquid.
 In (g), the right-hand-side angle is the only angle that is associated with a splay texture, hence $m_3=1$. The two on the left are associated with bend textures that give $m_4=2$. In total, using \eqref{Woutside}, we have $W=-1/2$, corresponding to a defect line in this case.

 The obstacle in plot (h) has an interesting packman shape. Viewed from the nematic liquid,
 a splay pattern can be found near
 the acute angle in the center, and bend patterns
near the two reflex angles on the right. This makes $m_1=1$ and $m_4=2$, therefore according to
\eqref{Woutside}, the total winding number of the defects is $W=-3/2$; indeed, there are three
$-1/2$ point defects in the system.

\bigskip
\section{Total Winding number of a nematic state of a complex geometry}\label{W3}

\subsection{Theory}

Finally, we generalize the above defect rules, \eqref{Winside} and \eqref{Woutside}, to the case of
a nematic liquid confined in a 2D boundary. Inside the nematic liquid, there are
 $M$ intruding obstacles of different shapes,
  forming different defect patterns nearby. These obstacles are labeled $l=1,...,M$.
The rule for the total winding number can be easily deduced based on \eqref{Winside} and \eqref{Woutside}. The sum of
the two gives
\begin{equation}\label{Wsum}
   W = -{1\over 2} \left\{(n_1- n_3-2) +  \sum_{l=1}^M [m_{1}^{(l)}- m_{3}^{(l)} +2]\right\},
\end{equation}
where $m_{1}^{(l)}$ and $m_{3}^{(l)}$ are the number of acute and reflex angles associated with splay patterns of the $l$th obstacle, respectively.

Then, it gives a final
\begin{equation}\label{Wall}
 W = -{1\over 2}(N_1-N_3 + 2M-2).
\end{equation}
Here, $N_1  $ is the total number of acute angles and $N_3  $ total number of reflex angles, all related to splay patterns nearby. In case of the occurrence of a splay pattern at a curved boundary, an extended angle is constructed from the tangent lines at the terminal ends of a curve.
The above rule could also be viewed as a general expression that contains both rules \eqref{Winside} and \eqref{Woutside}. For example, letting $M=0$ we return to \eqref{Winside}. Regarding the boundaryless texture in the far field of Sect.~\ref{W3} as having a hypothetical, inverted boundary that does not contribute to the defect pattern, we return to \eqref{Woutside} by letting $M=1+1=2$, where the additional $1$ takes the hypothetical boundary into account.

\subsection{Examples}

\begin{figure}[!t]
\centering
  \includegraphics[width=1\columnwidth]{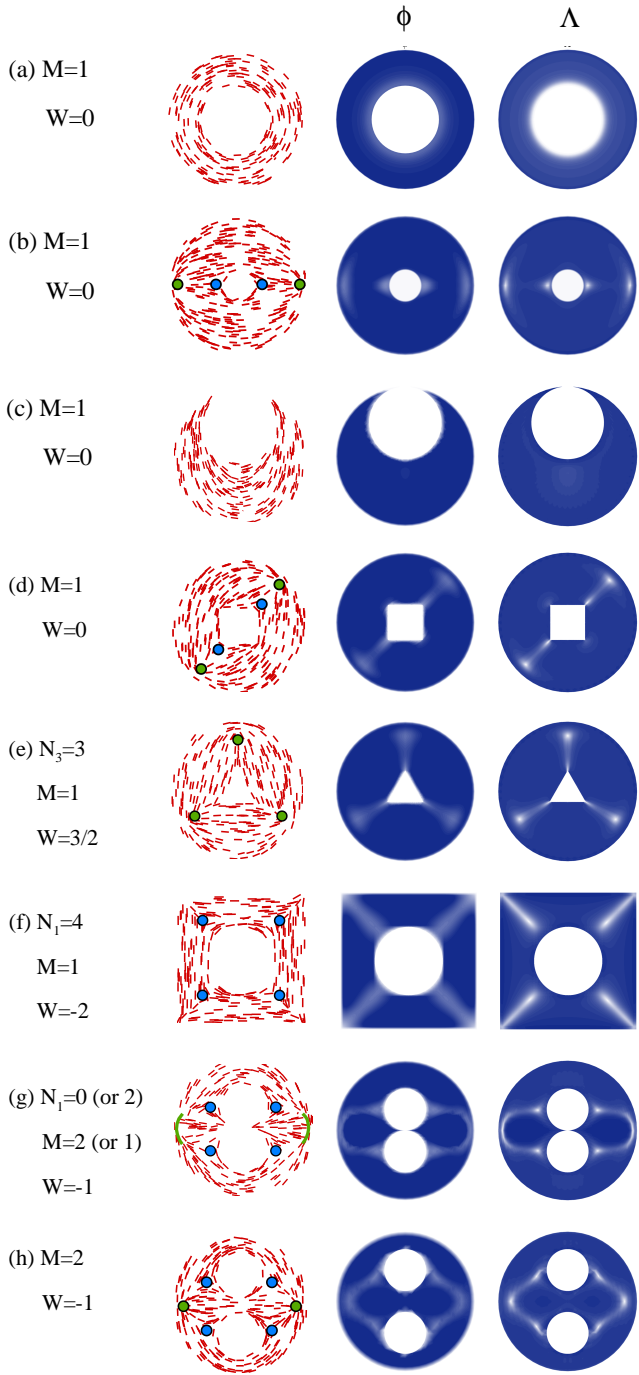}
  \caption{Examples of
defect patterns obtained from the solution to the extended Onsager model.
The reduced density used is $\tilde \rho = 10$ and $\tilde{L}$ is selected in the range
$[0.1, 0.2]$.
Colors used to indicate the defect points
and gray scale used for the density plots are the same as those
 in Fig. \ref{FIG3}. The non-zero values of $N_1, N_3$ and $M$, together with the total $W$, are specified in each plot.
    \label{FIG5}}
\end{figure}

A number of examples from solving the extended Onsager model are shown in Fig. \ref{FIG5}, mixing circular, triangle, and  square boundary conditions in various forms. The total winding number in these examples are compared with the general theoretical prediction in \eqref{Wall}.

The case of  annularly confined liquid crystals is a recent topic of significant interest \cite{Garlea2016Finite,Yao2020,Aarts2021Particle} [see Figs.~\ref{FIG5}(a) and (b)]. For homogeneous boundary conditions, the wall boundaries enforce the liquid-crystal
molecules to align along the wall direction, forming bend texture only. Hence $N_1=N_3=0$. Taking $M=1$ for a single obstacle, according  \eqref{Wall}, $W=0$. The liquid crystal texture is then either defect-free [as in (a)]
or having all defect winding numbers canceling each other [as in (b)].
While co-centered boundaries are shown in Figs.~\ref{FIG5}(a) and (b), the above rule is also true for non-co-centered cases.
The boundaries in (c) could be viewed as two non-co-centered circles asymptotically close to tangent contact, hence $N_1=N_3=0$, $M=1$, which gives $W=0$. They could also be viewed as forming a single boundary, for which we return to
the theory in Sect. \ref{W1}, where two splay patterns exist ($n_1=2$), hence
$W=0$. Both analyses give the same $W$.

Though the center obstacle in Fig.~\ref{FIG5}(d) is a square, the liquid crystal pattern around all four corners is bend. According to \eqref{Wall}, as it accounts for splay-related angles only, for $M=1$ we have $W=0$. The two pairs of $\pm 1/2$ defects in (d) cancel each other. From the perspective of having no splay-related angles, (d) is in the
same class as (a)-(c).

A contrasting case is Fig.~\ref{FIG5}(e), in which each corner of the central triangle
accompanies a splay pattern, hence $N_3=3$. With $M=1$, the defect rule \eqref{Wall}
gives $W=3/2$. The defect pattern in (e) clearly shows three +1/2 defect points.

Moving now to a case where the liquid crystal is confined in a square and excluded from a small, co-centered
circle [see Fig.~\ref{FIG5}(f)]. Here, the four splay patterns near the
square corners make $N_1 =4$ and the central circle does not contribute to
winding number counting. As the result, with $M=1$ we have $W=-2$,
which is the sum of the four $-1/2$ defects diagonally located inside the square.

The geometry in Fig.~\ref{FIG5}(g) can be assessed
by two different methods. In the first one, it can be regarded as the same as
the one in (h), with two inner circles separated from each other, but in this case, closely spaced. We then have
$N_1=N_3=0$ and $M=2$, as can be clearly identified from (h). The defect rule
 \eqref{Wall} then gives $W=-1$. On the other hand,
 the geometry in (g) can also be regarded as having one single obstacle ($M=1$) which contains
 two splay-related acute angles ($N_1=2$). Either method makes $W=-1$, which is consistent with the properties of the six defect points and lines in (g).

\begin{center}
\begin{table*}[!t]
\caption{Comparison between the defect rule, Eq. \eqref{Wall}, and selected defect patterns found in the recent literature.
The third column contains the figure numbers in the original references.
$M$ is the number of enclosed obstacles, $N_1$ and $N_3$ are splay related angles or extended angles if
smooth curves are involved. $W$ is the total winding number calculated from Eq. \eqref{Wall} and it matches the summed winding numbers of defects in the original figures.
\label{table}}\label{Tab1}
\begin{tabular}{|l|l|l|l||c|c|c|c|}
  \hline  \hline
Reference                                          & Approach         & Figure number        & Confinement type       & $M$   & $N_1$ & $N_3$ & $W$ \\
  \hline
Dzubiella2000 \citen{Dzubiella2000Topological}     & MC               & 16                   & Circle                 & 0     & 0     & 0     & 1\\
Galanis2006 \citen{Galanis2006Spontaneous}         & Experiment       & 4(a)                 & Square                 & 0     & 2     & 0     & 0\\
                                                   &                  & 4(b, c)              & Circle                 & 0     & 2     & 0     & 0\\
Tsakonas2007 \citen{Tsakonas2007}                  & LdG              & 3                    & Square                 & 0     & 2     & 0     & 0\\
Galanis2010 \citen{Galanis2010}                    & Experiment       & 1(a)                 & Circle                 & 0     & 0     & 0     & 1\\
Soares~e~Silva2011 \citen{Mulder2011}              & Experiment       & 2                    & Square                 & 0     & 4     & 0     & -1\\
Luo2012 \citen{Luo2012}                            & LdG              & 1                    & Square                 & 0     & 2     & 0     & 0\\
Chen2013 \citen{Chen2013}                          & Extended Onsager & 1                    & Square                 & 0     & 4     & 0     & -1\\
                                                   &                  & 2                    & Circle                 & 0     & 0     & 0     & 1\\
Lewis2014 \citen{Lewis2014}                        & OF               & 1                    & Rectangle              & 0     & 2     & 0     & 0\\
                                                   & Experiment       & 5($D, U_1$)          & Rectangle              & 0     & 2     & 0     & 0\\
                                                   & Experiment       & 5($D^*, U_1^*$)      & Rectangle              & 0     & 4     & 0     & -1\\
de~las~Heras2014 \citen{Heras2014Domain}           & MC               & 2(b, c)              & Circle                 & 0     & 0     & 0     & 1\\
Geigenfein2015 \citen{Geigenfein2015}              & MC               & 12(b)                & Square                 & 0     & 4     & 0     & -1\\
G{\^ a}rlea2015 \citen{Mulder2015}                 & MC               & 2(b)                 & Square                 & 0     & 4     & 0     & -1\\

G{\^ a}rlea2016 \citen{Garlea2016Finite}           & MC               & 1(e)                 & Circle                 & 0     & 0     & 0     & 1\\
                                                   &                  & 1(g)                 & Circle                 & 0     & 2     & 0     & 0\\
                                                   &                  & 2(a-c)               & Annulus                & 1     & 0     & 0     & 0\\
                                                   & Experiment       & 4(a)                 & Circle                 & 0     & 0     & 0     & 1\\
                                                   &                  & 4(b)                 & Annulus                & 1     & 0     & 3     & -3/2\\
Everts2016 \citen{Everts2016A}                     & LdG              & 6, 7(e-l)            & Square                 & 0     & 4     & 0     & -1\\
Robinson2017 \citen{Majumdar2017}                  & MC               & 2(b)right, 3 right, 4, 6   & Square           & 0     & 4     & 0     & -1\\
                                                   &                  & 5(a)                 & Square                 & 0     & 2     & 0     & 0\\
                                                   & LdG              & 11(1-5, 10-12, 14, 15)  & Square                 & 0     & 2     & 0     & 0\\
                                                   &                  & 11(6)                & Square                 & 0     & 0     & 0     & 1\\
                                                   &                  & 11(7)                & Square                 & 0     & 4     & 0     & -1\\
                                                   &                  & 11(8)                & Square                 & 0     & 3     & 0     & -1/2\\
                                                   &                  & 11(9, 13)             & Square                 & 0     & 1     & 0     & 1/2\\
Cortes2017 \citen{Cortes2017}                      & Experiment       & 3(N)                 & Square                 & 0     & 4     & 0     & -1\\
Yao2018 \citen{Yao2018}                            & Extended Onsager & 3, 4, 7              & Rectangle              & 0     & 4     & 0     & -1\\
Wang2018 \citen{Wang2018Order}                     & LdG              & 17(h)                & Square-in-square       & 1     & 1     & 1     & 0\\
                                                   &                  & 17(m)                & Square-in-square       & 1     & 2     & 3     & 1/2\\
                                                   &                  & 17(q)                & Square-in-square       & 1     & 2     & 1     & -1/2\\
Garlea2019 \citen{Garlea2019Colloidal}             & MC               & 2(except H)          & Circle and lens-shape  & 0     & 0     & 0     & 1\\
                                                   & Experiment       & 2(except H)          & Circle and lens-shape  & 0     & 0     & 0     & 1\\
Hashemi2019 \citen{Hashemi2019Finding}             & MC               & 1(b-d), 3, 5(b-d)    & Circle                 & 0     & 0     & 0     & 1\\
Hashemi2019 \citen{Hashemi2019Structure}           & MC               & 1(b, c), 2(b, c), 3(c), 7& Square               & 0     & 4     & 0     & -1\\
Yin2020 \citen{Yin2020Construction}                & LdG              & 2(g) C$\pm$, I$\pm$  & Square                 & 0     & 0     & 0     & 1\\
                                                   &                  & 2(g) S, H, I, D      & Square                 & 0     & 2     & 0     & 0\\
                                                   &                  & 2(g)T                & Square                 & 0     & 3     & 0     & -1/2\\
Han2020 \citen{Han2020A}                           & LdG              & 4                    & Hexagon                & 0     & 0     & 0     & 1\\
                                                   & LdG              & 9                    & Hexagon                & 0     & 2     & 0     & 0\\
Han2021 \citen{Han2021Solution}                    & LdG              & 2(a, b)              & Hexagon                & 0     & 0     & 0     & 1\\
                                                   &                  & 2(c, d)              & Hexagon                & 0     & 2     & 0     & 0\\
                                                   &                  & 5(c)                 & Triangle               & 0     & 3     & 0     & -1/2\\
  \hline
  \hline
\end{tabular}\label{table1}
\end{table*}
\end{center}

\bigskip
\section{Discussion}
\subsection{Comparison with the literature}\label{Comp}

In the above, the defect rules of a confined nematic liquid  are discussed in light of solutions to the extended Onsager model as examples. The rules are quite general, independent of the actual theoretical or experimental approaches used in studying confined nematic liquids, as long as the liquid crystal molecules near a confinement wall
prefer to align in parallel with the wall surface.
Table \ref{table1} contains an incomplete list of some examples
found in the literatures and the comparison with our defect rules.

In this list, the type of the actual liquid-crystal ``molecules'' varies in a large range.
The images of defect patterns were directly observed by crossed polarizers
 on
liquid crystal E7, confined in square cells, which was then compared to a LdG theory\cite{Tsakonas2007}.
The observation
of defect patterns of well-equilibrated, real macroscopic steel needles confined in square and circular cells was also made and compared to the solution of an OF theory \cite{Galanis2006Spontaneous,Galanis2010}.
Although biomolecules are usually characterized by their semiflexibility,
confined in finite geometries, they also show nematic textures, some containing defect patterns \cite{Lewis2014,Garlea2016Finite,Garlea2019Colloidal}.
A confocal-microscopy image of the nematic layer of micron-size
rodlike colloid particles in a square well has also shown a defect pattern that contains defects
\cite{Cortes2017}.

The list also includes
the defect patterns produced from
 a number of
theoretical approaches taken in studying the confined liquid-crystal systems.
A short-cut to study the orientationally ordered state is the use of
a model similar to the original OF theory \cite{ericksen1991liquid,deG1993}. Typically, the orientational properties are over-simplified by using a main-axis director field only, which is a unit vector field depending on the spatial location $\mathbf{r}$; the free-energy is then proposed in terms of spatial derivatives of the vector field, where, at this stage, some of the anticipated orientational-ordering properties are taken into account.
This has been one of the popular approaches to describe mechanical distortions (bend, splay, twist, etc.) of the director field in response to the external force. For confined liquid crystals, for example, OF theories have been used to explain experimental observations\cite{Galanis2006Spontaneous,Lewis2014,Galanis2010}.

The LdG theory for a system composed of rodlike molecules
calls for the identification of a second-order, $3\times 3$ tensor order parameter, in which elements are functions of $\mathbf{r}$.
The LdG theory contains physical parameters
associated with the elastic energy, typically depending on the molecular structure. A commonly used approximation is the one-coefficient approach, which erases the molecular identity and ignores, e.g., molecular flexibility of a molecule. The concept of the director field is not used in LdG originally and, instead, is produced as a result of the model. Incorporating the Dirichlet boundary conditions that enforces parallel alignment of the nematic directors at the confinement boundary, this has been a popular approach in recent studies of the liquid confinement problems \cite{Luo2012,Majumdar2017,Wang2018Order,Han2020A,Han2021Solution}.

A density functional theory (DFT) focuses on the probability distribution which is an inhomogeneous function of
molecular orientation described by the unit vector $\bu$ and molecular spatial arrangement described by $\mathbf{r}$. Various forms of have been used for the liquid-crystal confinement problem, with incorporation of boundary conditions.
The extended Onsager model , for example, neglects the free-energy expansion beyond the second virial level, contains sufficient orientation-orientation interaction that describes the nematic state \cite{Chen2013,Lewis2014,Yao2018,Yao2020}. Built in a more complex form, the DFTs can effectively include high order virial terms and have a tool for studying liquid crystals, in particular, here for 2D confinement in
Ref. \citen{Aarts2021Particle}.

Beyond experiential and theoretical approaches,
direct computer simulations of liquid crystal molecules in confinement have also been taken. Typically,
a liquid of anisotropically shaped molecules are placed in a confinement box; their positions and orientations
are updated either according to the molecular dynamics or the Monte Carlo (MC) transition probability. Then, either snapshots or overall statistics can be collected. For a sensible comparison, only those
configurations that follow the parallel homogeneous boundary patterns are included in this table
\cite{Dzubiella2000Topological,Heras2014Domain,Mulder2015,Garlea2016Finite,Garlea2019Colloidal,Hashemi2019Structure,Hashemi2019Finding}.

As can be summarized in Table \ref{table1}, a rich
variety of liquid crystal defects have been produced either experimentally or theoretically.
Regardless of the actual experimental systems and the theoretical approaches taken to study the problem of confinement liquid crystals, all the defect patterns observed can be accounted for by our defect rules.

\subsection{Other methods of counting winding numbers}

The total winding number $W$ considered in this paper is based on the method of
 taking a calculation loop around the wall boundary and summing up the winding number of every enclosed defects inside the loop. There are other methods of defining the total winding number.

In Refs. \citen{Majumdar2017,Han2021Solution},
for example, for a liquid crystal confined in a regular polygon of $n$ sides, an addition contribution from every polygon
corner is added to $W$.
For square confinement, $n=4$, \citeauthor{Majumdar2017} added a winding number $1/2-1/n = 1/4$ from
a splay-associated corner and $-1/n=-1/4$ from a bend-associated corner to $W$. This makes the total winding number to be\cite{Majumdar2017}
\begin{equation}\label{WMajumdar}
W^\prime = W +
n_1\left({1\over 2}-{1\over n}\right) - {n_2\over n} =0,
\end{equation}
where $W= - (n_1 -2)/2$ from \eqref{Winside} for $n_3=0$ is inserted to the above.
The same $W^\prime=0$ was also used to explain the defect points observed
by \citeauthor{Han2021Solution} for hexagonal confinement $(n=6)$. Now
$1/2-1/n = 1/3$ from
a splay-associated corner and $-1/n = -1/6$ from a bend-associated corner
\cite{Han2021Solution}.
One could generalize this method even further for nonregular polygon confinement, which
would always gives a universal $W^\prime = 0$.

To explain the defect patterns displayed from the MC
simulations of hard ellipses confined by square boundary condition,
\citeauthor{Hashemi2019Structure} added four +1/2 winding
numbers to $W$, each from a square corner.
For four splay-associated corners, this makes $W^{\prime\prime} = W + 4/2 = 1$, which is
a value quoted in Ref. \citen{Hashemi2019Structure}. The reason for using such a $W^{\prime\prime}$ is
unknown.

\subsection{Splay or bend}\label{SplayBend}

\begin{figure}[!t]
\centering
  \includegraphics[width=1\columnwidth]{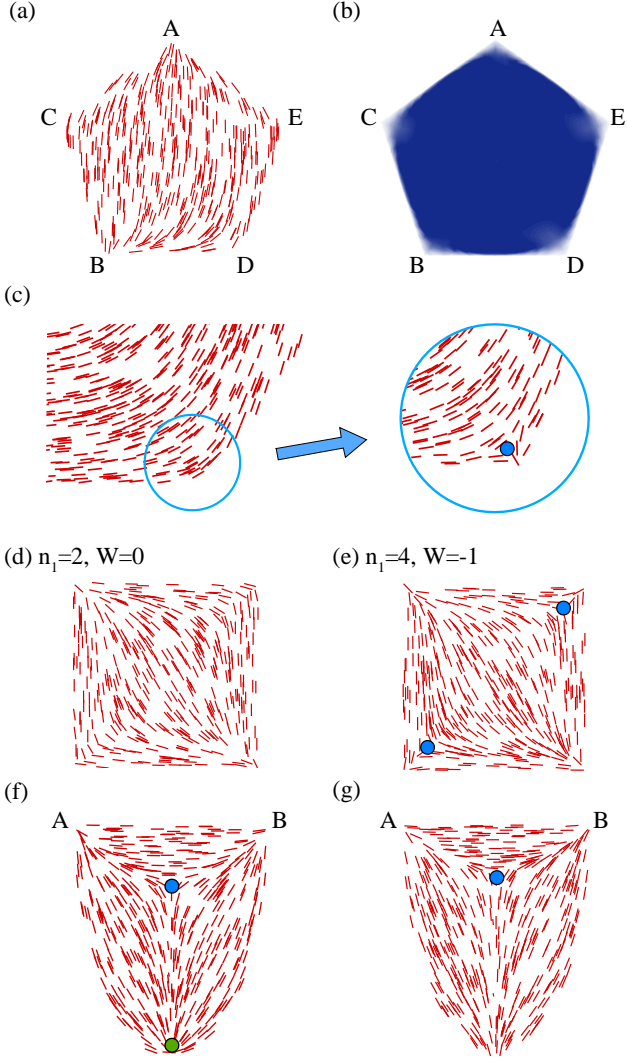}
  \caption{
  Examples used for the discussion of splay and bend nematic textures.
  Plots (a) and (b)
  are the nematic-director map and the density profile, respectively, for a pentagon-confined rodlike liquid that has no interior defect.
  Plot  (c) shows an example of
  a bend pattern in the zoom-out (coarse-grained) version, which is actually composed of a defect point
   and a splay pattern in the zoom-in (fine-grained) version.
   Plots (d) and (e) also demonstrates that two different views can be taken
   to view the bend-associated angles in (d). In (d), the two bend angles together with two splay-associated angles give an overall $W=0$. In (e), all corners are associated with splay patterns, which gives $W=-1$.
   Plots (f) and (g) are the nematic-director map of a nematic fluid confined in a shield-shape boundary, which have different
  degrees of resolution. In the fine-grained picture (f), a defect is visible at the bottom and in the coarse-grained
  picture (g), such a defect is invisible.
   }
    \label{FIG6}
\end{figure}


The theory on the total winding number developed in Sects. \ref{W1}-\ref{W3} depends on the identification of the number of
acute and reflex angles associated with a splay pattern in the nematic fluid. Those with a bend pattern are not taken into account
in the formulae. The classification of a bend pattern, however, deserves more detailed discussion.

The formation of the splay or bend patterns does not uniquely depend on the angle size. Although,
usually the liquid crystals inside a sharp acute angle display a splay pattern.
An example is given in Fig. \ref{FIG6}(a) where the molecules are confined in a regular pentagon.
Two and three angles
are associated with splay and bend textures respectively,
regardless of the fact that all angles by themselves are equivalent.
Near angles A and B, the molecules prefer to fill the near-angle space in a splay pattern, to avoid
density depletion near the angles, which would cost unwanted local depletion entropy. The density plot in (b)
clearly
demonstrates this fact near these angles.
Near
angles C, D, and E, the bend textures are clearly visible. The molecules near these angles would need to
make a compromise to accommodate
the bend texture, by leaving the immediate area inside the angles unfilled.
This affects the length scale of a several $L$, which is visible in the density plot. Although the
density depletion costs the entopic term in the local free energies,
the overall pattern over the entire fluid now has no defects, which is preferred by the total
system free energy. Hence, the formation of
the splay or bend patterns near an angle is completely determined by the balanced consideration of the total free energy.
In Ref. \citen{Han2021Solution}, Han et al. deemed that solutions with bend-like vertices had a higher Morse index than those with splay-like vertices when they investigated the solution landscape of nematic liquid crystals on a hexagon. Morse index, which implies the unstable directions that may affect the state, plays an important role in the solution landscape. As we know, the solution landscape is determined by the energy landscape, which further verifies that the vertex profile (splay-like or bend-like) has a lot to do with the total free energy.

Depending on the coarse-graining level,
the classification of a splay- or bend-associated angle is not unique. Sometimes two different views can be taken.
The example in Fig. \ref{FIG6}(c) shows a bend pattern at the acute angle at a scale much greater than the molecular length, hence the angle is not accounted for in the defect formula.
A second view gives a different accounting system, after observing at a molecular scale that the
acute angle is actually associated with a splay pattern, which connects with
a $-1/2$ defect point nearby. Using \eqref{Winside}, an additional $-1/2$ is produced by adding an extra-$1$ to $n_1$ because of
the splay-associated acute angle, but this addition $-1/2$ is completely used to describe the
$-1/2$ defect point close to the angle. Therefore, there are two ways of assessing this angle: either ignore the defect in the zoom-out version by regarding this angle as a bend-textured angle, or in a zoom-in version by accounting for the defect and the splay-associated angle. Both gives the correct total winding number counting.

Hence, a bend-associated acute angle can always be treated as a splay-associated acute angle with a hidden $-1/2$ defect point. If one takes this view, then $n_1$ in \eqref{Winside} could be regarded as the total number of acute angles of an confinement, and one simply has $W = -n_1/2 +1$.
On this basis, a seemingly bend-associated angle would need to be augmented by an invisible $-1/2$ defect point in using
the above. It is this view that was taken in Sect. II(D) of Ref.
 \citen{Yao2018}. For liquid crystal confined by a square boundary, examples in Figs. \ref{FIG6}(d) and
 (e) further demonstrates the concept, which gives $W=0$ (with two bend-associated angles)
 and $W=-1$ (with all splay-associated angles). Indeed, the confocal image of actin filaments in square confinement observed
 in Ref. \citen{Mulder2011} display the texture similar to Fig. \ref{FIG6}(d).

The interplay between splay and bend patterns can also manifest in another form.
The example in Fig. \ref{FIG6}(f) clearly has two splay-associated angles A and B hence $n_1=2$. Again,
two different views can be taken for consideration of the defect near the bottom, curved boundary.
The first view is to ignore the $+1/2$ defect point near the bottom curve,
shown in Fig. \ref{FIG6}(f) by the green circle. The entire pattern
then looks like Fig. \ref{FIG6}(g).
According to the theory presented in Sect. \ref{W1}, now the curved segment has an extended tangent angle that is associated with a splay pattern, hence it adds to the $n_1$ an additional $1$: $n_1=2+1=3$. The formula in Eq. \eqref{Winside}
then gives $W=-1/2$, which is fully consistent with the only defect point shown in Fig. \ref{FIG6}(g).
The second view gives a different accounting system, as illustrated in Fig. \ref{FIG6}(f).
The nematic fluid near the bottom curve has no singularity along the curved boundary, which produces an overall $W=0$
for the interior defects, according to \eqref{Winside}.
Indeed, the sum of
the winding number of the $+1/2$ defect (green circle in the figure) and the $-1/2$ defect (blue
circle) gives $W=0$.
Both views are consistent in winding number analysis. The fact that a $+1/2$ defect very close
to a smooth bend curve can be regarded as a splay-associated, extended angle
produces a
simple method to account for the winding number: one can take all curves as bend curves and then associate a hidden $+1/2$ defect with a splay defect at the curved boundary. This method, the same as the second view presented for the example, was taken in Sect. II(D) of Ref.
 \citen{Yao2018}.

\subsection{Extreme confinement}

When the ratio between the molecular length and  typical length-scale of confinement boundaries,
$L/a$, exceeds a critical value,  packing rodlike molecules in a finite confinement space dominates over the need to maximize the orientational entropy.
A characteristic property of these extreme confinement systems is
that
the nematic directors along the confinement boundary no longer prefer parallel alignment. This was already
observed in  earlier granular-particle experiments \cite{Galanis2006Spontaneous,Galanis2010}, recent ${\it fd}$-virus packing experiments \cite{Garlea2016Finite,Garlea2019Colloidal} and direct images of micron-sized colloidal particles \cite{Aarts2021Particle}. One the theoretical side,
Monte Carlo simulations \cite{Garlea2016Finite,Garlea2019Colloidal}
and the numerical solution to the extended Onsager model \cite{Yao2020} have both
demonstrated that these
extremely confined liquid crystals are ``thermodynamically'' stable extreme phases.
Furthermore, concrete evidences \cite{Galanis2006Spontaneous}, in particular a study of the free energy
\cite{Yao2020},
 all indicate that a phase transition exists between a usual confined state and an extreme state.
It is unclear how the LdG theory, which typically requires a Dirichlet boundary condition, can be applied to model an extreme state.

The destruction of the homogeneous boundary condition precludes
 the basic assumption used in
setting up
 the defect rules.
The total winding number formulae developed in this work are based on the assumption of a continuous, space-filling nematic
fluid, which is not applicable to the extreme nematic states.

\bigskip
\section{SUMMARY}

Summing up all individual winding numbers of defect points and lines in a two-dimensional, confined nematic liquid crystal, how does it
relate to the confining geometry formed by angles and curved segments? In this study, we deduced a
general defect rule, which is applicable to a nematic liquid crystal having homogeneous boundary conditions. As we demonstrated above, the determinant factor is the number of splay-related angles and curved segments,
whereas the bend-associated angles and curved segments do not contribute to the final result.

The general defect rule, \eqref{Wall}, which includes the special cases in \eqref{Winside} and \eqref{Woutside},
was then further validated by a comparison with results produced from experimental and theoretical studies
in Sect. \ref{Comp}.
While most of these studies concern systems composed of circular and polygon shapes,
additional confinement types were also supplemented by considering the numerical solutions to the extended Onsager model, in Sects~\ref{W1}, \ref{W2}, and \ref{W3}.

The main focus of the current study is specifically on 2D liquid crystals confined by a closed line boundary. 
It fits into the much greater scope of the general topic of liquid crystals in confinement.
Within the general topic, 
a well-established theorem is for liquid crystals confined on a curved and closed surface, such as on the spherical surface or colloidal surface, for which
the total winding number formed by the defects in
the liquid-crystal director lines is dictated by the 
Euler characteristics of the surface.
Also within the general topic but for the specific confinement type of liquid crystals,  
on a flat surface and enclosed by a line boundary, the theory 
established in this paper is complementary to this theorem and consistently explains the 
variety of defect patterns observed in  the recent literature.

\section{\bf ACKNOWLEDGEMENTS}

We wish to acknowledge the financial support from the National Natural Science Foundation of China (Grant No. 21873009 and No. 12050002) and the Natural Sciences and Engineering Council of Canada.

\appendix

\section{Extended Onsager model}\label{A1}

Assume that the distribution density function of finding the centers-of-mass of rodlike molecules at a spatial position specified by the vector $\br$ with the condition that the rods point at the direction specified by the unit vector $\bu$ is
$\rho_c(\br,\bu)$. It normalizes to $n$, the number of confined rodlike molecules in an area $A$,
 $\intn\dd \br\intn\dd \bu \rho_c(\br,\bu) =n $. Accurate to the second-virial term \cite{pathriastatistical}, the free energy of the system can be written in a truncated Mayer expansion
\begin{equation}
\begin{aligned}\label{Fdef}
\beta &F=\intn\rho_c(\mathbf{r},\mathbf{u}) \ln [L^2\rho_c(\mathbf{r},\mathbf{u}) ] {\rm d}\mathbf{r} {\rm d}\mathbf{u} + \intn\rho_c(\mathbf{r},\mathbf{u}) V_c(\mathbf{r},\mathbf{u})  {\rm d}\mathbf{r} {\rm d}\mathbf{u}
 \\
&+ \frac{1}{2}\int\rho_c(\mathbf{r},\mathbf{u}) w(\mathbf{r},\mathbf{u};\mathbf{r}^\prime, \mathbf{u}^\prime)\rho_c(\mathbf{r}^\prime, \mathbf{u}^\prime) \;\dd\mathbf{r} \dd\mathbf{u} \dd\mathbf{r}^\prime \dd\mathbf{u}^\prime,
\end{aligned}
\end{equation}
where $\beta = 1/k_B T$, with $k_B$ being the Boltzmann constant and $T$ the temperature.
The first term represents the entropy of a spatially inhomogeneous and orientationally ordered fluid of rodlike molecules, where $L^2$ is included for dimensional convenience.
The third term takes into account the interaction between two
rodlike molecules having the coordinates $(\mathbf{r},\mathbf{u})$ and $(\mathbf{r}^\prime, \mathbf{u}^\prime)$, where
the Mayer function  $-w = \exp (-\beta v)-1$ . The interaction potential energy $v$ between the two rigid molecules takes a value $v=\infty$
when the configuration of two rods overlap; $v=0$ otherwise.
The vector $\mathbf{u}$ are represented by the variables $\theta$, the angle a rodlike molecule makes with respect to the horizontal axis.

The second term describes the interaction between a single rodlike molecule with an external potential energy. In the current application, $V_c =0$ if the rodlike molecule has no overlap with a boundary wall, and $V_c =\infty$ if it does. Unlike the wall-potential for a small molecule where the orientation is not a concern, the rod-wall interaction depends on the orientation $\mathbf u$. In the numerical calculation, we used $V_c =10^3$ instead of $\infty$; this effectively produces; this effectively
produces $\rho_c<0.005$ when part of a rod overlaps with the wall. This masking technique is computationally efficient and requires no explicit specification of the boundary condition of $\rho_c(\mathbf{r},\mathbf{u})$. The expense, of course, is the need to careful specify $V_c(\mathbf{r},\mathbf{u})$ for a particular confinement shape.

In a much simpler mathematical problem, \citeauthor{Onsager1949} considered a trial-function solution of the model for a spatially homogeneous system (with $V_c=0$) where $\rho_c(  \mathbf{r},\mathbf{u})$ is a function of $\bu$ only to demonstrate the existence of the nematic phase\cite{Onsager1949}.
In 2D, one can take a bifurcation analysis and show that the  second-order isotropic-nematic phase transition takes place when the 2D
particle density $\rho_0= n/A$ reaches a critical $\rho_0^* L^2 = 3\pi/2$ \cite{Kayser1978,Cuesta1989,Chen1993prl}. Most of $\tilde \rho$ values used here are well-above this critical density.

The reduced free energy in \eqref{Fdef} is the extended version of the Onsager model and contains $\mathbf r$-dependence.
As a functional of the function $\rho_c(\mathbf{r},\mathbf{u})$, it needs to be minimized, by solving the stationary condition,
\begin{equation}\label{DFdr}
{\delta F\over \delta \rho_c(\mathbf{r},\mathbf{u})} =0.
\end{equation}
The actual calculation is conducted by mapping the current problem
to the equivalent self-consistent field theory of a wormlike-chain system, where the chain rigidity is taken to be infinity
\cite{Chen2013,Chen2016}.
The current numerical scheme used in solving the
 the Green's formalism of the problem is identical to the procedure documented in an Appendix of Ref. \citen{Yao2018}, with the addition of an external energy as the masking potential to mimic the boundary condition.

\section{Visualization of the structures}

In the text, a number of physical properties are analyzed and displayed, calculated from the distribution function of the center-of-mass of a rodlike molecule, $\rho_c(\br, \theta)$, obtained from minimizing the free energy.
One can deduce the distribution density function for segments on the rodlike molecules, regardless the position on the rod, by defining
\begin{equation}
f(\br,\bu)=\frac{1}{\rho_0}\int_0^1\rho_c\left[\br-\bu L\left(s-\frac{1}{2}\right),\bu\right]\dd s,
\end{equation}
where the distribution of the segments at the path coordinate $s$ is traced back to the rod centers.
The integrant represents the probability density of finding the segment labeled by $s$ on the rodlike molecule to appear at a location with the coordinate $\br$.
With this definition, $f(\br,\bu)$ is dimensionless. 


A number of properties are calculated by using $f(\br,\bu)$. The distribution density function for rod segments is calculated from
\begin{equation}
\phi(\br)=\int_0^{2\pi}f(\br,\theta)\dd\theta,
\end{equation}
which is plotted Figs. \ref{FIG3},\ref{FIG4}, \ref{FIG5} and \ref{FIG6}(b). The $2\times 2$ $\sf Q$-tensor,
\begin{equation}  \label{Qdef}
  \begin{split}
    {\sf Q}(\br) &= {1\over 2} \left[\begin{array}{cc}
      S(\br) & T(\br)\\
      T(\br) & -S(\br)
    \end{array}\right],\\
  \end{split}
\end{equation}
is calculated from
\begin{align}
S(\br)&=\frac{\int_0^{2\pi}\dd\theta\cos(2\theta)f(\br,\theta)}{\phi(\br)},\\
T(\br)&=\frac{\int_0^{2\pi}\dd\theta\sin(2\theta)f(\br,\theta)}{\phi(\br)}.
\end{align}
Both $S$ and $T$ characterize the orientational ordering of the rodlike molecules by themselves and can be used directly. The scalar orientational order parameter is determined by the positive eigenvalue of the $\sf Q$-tensor,
\begin{equation}\label{lambda}
\Lambda(\br)= \sqrt{S^2(\br)+T^2(\br)},
\end{equation}
which is plotted in Figs. \ref{FIG3}, \ref{FIG4}, and \ref{FIG5}.
Particularly, the locations where $\Lambda \to 0$ are considered as defect points.




\bibliography{Yao3} 

\begin{thebibliography}{78}
\expandafter\ifx\csname natexlab\endcsname\relax\def\natexlab#1{#1}\fi
\expandafter\ifx\csname bibnamefont\endcsname\relax
  \def\bibnamefont#1{#1}\fi
\expandafter\ifx\csname bibfnamefont\endcsname\relax
  \def\bibfnamefont#1{#1}\fi
\expandafter\ifx\csname citenamefont\endcsname\relax
  \def\citenamefont#1{#1}\fi
\expandafter\ifx\csname url\endcsname\relax
  \def\url#1{\texttt{#1}}\fi
\expandafter\ifx\csname urlprefix\endcsname\relax\def\urlprefix{URL }\fi
\providecommand{\bibinfo}[2]{#2}
\providecommand{\eprint}[2][]{\url{#2}}

\bibitem[{\citenamefont{de~Gennes and Prost}(1993)}]{deG1993}
\bibinfo{author}{\bibfnamefont{P.~G.} \bibnamefont{de~Gennes}}
  \bibnamefont{and} \bibinfo{author}{\bibfnamefont{J.}~\bibnamefont{Prost}},
  \emph{\bibinfo{title}{The Physics of Liquid Crystals}}
  (\bibinfo{publisher}{Clarendon Press}, \bibinfo{address}{Oxford},
  \bibinfo{year}{1993}).

\bibitem[{\citenamefont{Poulin et~al.}(1997)\citenamefont{Poulin, Stark,
  Lubensky, and Weitz}}]{Poulin1997}
\bibinfo{author}{\bibfnamefont{P.}~\bibnamefont{Poulin}},
  \bibinfo{author}{\bibfnamefont{H.}~\bibnamefont{Stark}},
  \bibinfo{author}{\bibfnamefont{T.~C.} \bibnamefont{Lubensky}},
  \bibnamefont{and} \bibinfo{author}{\bibfnamefont{D.~A.} \bibnamefont{Weitz}},
  \bibinfo{journal}{Science} \textbf{\bibinfo{volume}{275}},
  \bibinfo{pages}{1770} (\bibinfo{year}{1997}),
  \urlprefix\url{https://doi.org/10.1126/science.275.5307.1770}.

\bibitem[{\citenamefont{Nazarenko et~al.}(2001)\citenamefont{Nazarenko, Nych,
  and Lev}}]{Nazarenko2001}
\bibinfo{author}{\bibfnamefont{V.~G.} \bibnamefont{Nazarenko}},
  \bibinfo{author}{\bibfnamefont{A.~B.} \bibnamefont{Nych}}, \bibnamefont{and}
  \bibinfo{author}{\bibfnamefont{B.~I.} \bibnamefont{Lev}},
  \bibinfo{journal}{Phys. Rev. Lett.} \textbf{\bibinfo{volume}{87}},
  \bibinfo{pages}{075504} (\bibinfo{year}{2001}),
  \urlprefix\url{https://doi.org/10.1103/PhysRevLett.87.075504}.

\bibitem[{\citenamefont{Mu{\v{s}}evi{\v{c}}
  et~al.}(2006)\citenamefont{Mu{\v{s}}evi{\v{c}}, {\v{S}}karabot, Tkalec,
  Ravnik, and {\v{Z}}umer}}]{Muvsevivc2006two}
\bibinfo{author}{\bibfnamefont{I.}~\bibnamefont{Mu{\v{s}}evi{\v{c}}}},
  \bibinfo{author}{\bibfnamefont{M.}~\bibnamefont{{\v{S}}karabot}},
  \bibinfo{author}{\bibfnamefont{U.}~\bibnamefont{Tkalec}},
  \bibinfo{author}{\bibfnamefont{M.}~\bibnamefont{Ravnik}}, \bibnamefont{and}
  \bibinfo{author}{\bibfnamefont{S.}~\bibnamefont{{\v{Z}}umer}},
  \bibinfo{journal}{Science} \textbf{\bibinfo{volume}{313}},
  \bibinfo{pages}{954} (\bibinfo{year}{2006}),
  \urlprefix\url{https://doi.org/10.1126/science.1129660}.

\bibitem[{\citenamefont{{\v{S}}karabot
  et~al.}(2007)\citenamefont{{\v{S}}karabot, Ravnik, {\v{Z}}umer, Tkalec,
  Poberaj, Babi{\v{c}}, Osterman, and Mu{\v{s}}evi{\v{c}}}}]{vskarabot2007two}
\bibinfo{author}{\bibfnamefont{M.}~\bibnamefont{{\v{S}}karabot}},
  \bibinfo{author}{\bibfnamefont{M.}~\bibnamefont{Ravnik}},
  \bibinfo{author}{\bibfnamefont{S.}~\bibnamefont{{\v{Z}}umer}},
  \bibinfo{author}{\bibfnamefont{U.}~\bibnamefont{Tkalec}},
  \bibinfo{author}{\bibfnamefont{I.}~\bibnamefont{Poberaj}},
  \bibinfo{author}{\bibfnamefont{D.}~\bibnamefont{Babi{\v{c}}}},
  \bibinfo{author}{\bibfnamefont{N.}~\bibnamefont{Osterman}}, \bibnamefont{and}
  \bibinfo{author}{\bibfnamefont{I.}~\bibnamefont{Mu{\v{s}}evi{\v{c}}}},
  \bibinfo{journal}{Phys. Rev. E} \textbf{\bibinfo{volume}{76}},
  \bibinfo{pages}{051406} (\bibinfo{year}{2007}),
  \urlprefix\url{https://doi.org/10.1103/PhysRevE.76.051406}.

\bibitem[{\citenamefont{Ognysta et~al.}(2008)\citenamefont{Ognysta, Nych,
  Nazarenko, Mu{\v{s}}evi{\v{c}}, {\v{S}}karabot, Ravnik, {\v{Z}}umer, Poberaj,
  and Babi{\v{c}}}}]{ognysta20082d}
\bibinfo{author}{\bibfnamefont{U.}~\bibnamefont{Ognysta}},
  \bibinfo{author}{\bibfnamefont{A.}~\bibnamefont{Nych}},
  \bibinfo{author}{\bibfnamefont{V.}~\bibnamefont{Nazarenko}},
  \bibinfo{author}{\bibfnamefont{I.}~\bibnamefont{Mu{\v{s}}evi{\v{c}}}},
  \bibinfo{author}{\bibfnamefont{M.}~\bibnamefont{{\v{S}}karabot}},
  \bibinfo{author}{\bibfnamefont{M.}~\bibnamefont{Ravnik}},
  \bibinfo{author}{\bibfnamefont{S.}~\bibnamefont{{\v{Z}}umer}},
  \bibinfo{author}{\bibfnamefont{I.}~\bibnamefont{Poberaj}}, \bibnamefont{and}
  \bibinfo{author}{\bibfnamefont{D.}~\bibnamefont{Babi{\v{c}}}},
  \bibinfo{journal}{Phys. Rev. Lett.} \textbf{\bibinfo{volume}{100}},
  \bibinfo{pages}{217803} (\bibinfo{year}{2008}),
  \urlprefix\url{https://doi.org/10.1103/PhysRevLett.100.217803}.

\bibitem[{\citenamefont{Ravnik et~al.}(2007)\citenamefont{Ravnik,
  {\v{S}}karabot, {\v{Z}}umer, Tkalec, Poberaj, Babi{\v{c}}, Osterman, and
  Mu{\v{s}}evi{\v{c}}}}]{Ravnik2007entangled}
\bibinfo{author}{\bibfnamefont{M.}~\bibnamefont{Ravnik}},
  \bibinfo{author}{\bibfnamefont{M.}~\bibnamefont{{\v{S}}karabot}},
  \bibinfo{author}{\bibfnamefont{S.}~\bibnamefont{{\v{Z}}umer}},
  \bibinfo{author}{\bibfnamefont{U.}~\bibnamefont{Tkalec}},
  \bibinfo{author}{\bibfnamefont{I.}~\bibnamefont{Poberaj}},
  \bibinfo{author}{\bibfnamefont{D.}~\bibnamefont{Babi{\v{c}}}},
  \bibinfo{author}{\bibfnamefont{N.}~\bibnamefont{Osterman}}, \bibnamefont{and}
  \bibinfo{author}{\bibfnamefont{I.}~\bibnamefont{Mu{\v{s}}evi{\v{c}}}},
  \bibinfo{journal}{Phys. Rev. Lett.} \textbf{\bibinfo{volume}{99}},
  \bibinfo{pages}{247801} (\bibinfo{year}{2007}),
  \urlprefix\url{https://doi.org/10.1103/PhysRevLett.99.247801}.

\bibitem[{\citenamefont{{\v{S}}karabot
  et~al.}(2008)\citenamefont{{\v{S}}karabot, Ravnik, {\v{Z}}umer, Tkalec,
  Poberaj, Babi{\v{c}}, Osterman, and
  Mu{\v{s}}evi{\v{c}}}}]{vskarabot2008interactions}
\bibinfo{author}{\bibfnamefont{M.}~\bibnamefont{{\v{S}}karabot}},
  \bibinfo{author}{\bibfnamefont{M.}~\bibnamefont{Ravnik}},
  \bibinfo{author}{\bibfnamefont{S.}~\bibnamefont{{\v{Z}}umer}},
  \bibinfo{author}{\bibfnamefont{U.}~\bibnamefont{Tkalec}},
  \bibinfo{author}{\bibfnamefont{I.}~\bibnamefont{Poberaj}},
  \bibinfo{author}{\bibfnamefont{D.}~\bibnamefont{Babi{\v{c}}}},
  \bibinfo{author}{\bibfnamefont{N.}~\bibnamefont{Osterman}}, \bibnamefont{and}
  \bibinfo{author}{\bibfnamefont{I.}~\bibnamefont{Mu{\v{s}}evi{\v{c}}}},
  \bibinfo{journal}{Phys. Rev. E} \textbf{\bibinfo{volume}{77}},
  \bibinfo{pages}{031705} (\bibinfo{year}{2008}),
  \urlprefix\url{https://doi.org/10.1103/PhysRevE.77.031705}.

\bibitem[{\citenamefont{Ravnik and {\v{Z}}umer}(2009)}]{Ravnik2009braids}
\bibinfo{author}{\bibfnamefont{M.}~\bibnamefont{Ravnik}} \bibnamefont{and}
  \bibinfo{author}{\bibfnamefont{S.}~\bibnamefont{{\v{Z}}umer}},
  \bibinfo{journal}{Soft Matter} \textbf{\bibinfo{volume}{5}},
  \bibinfo{pages}{4520} (\bibinfo{year}{2009}),
  \urlprefix\url{https://doi.org/10.1039/b913065a}.

\bibitem[{\citenamefont{Ognysta et~al.}(2011)\citenamefont{Ognysta, Nych,
  Uzunova, Pergamenschik, Nazarenko, \ifmmode~\check{S}\else \v{S}\fi{}karabot,
  and Mu\ifmmode \check{s}\else \v{s}\fi{}evi\ifmmode~\check{c}\else
  \v{c}\fi{}}}]{Ognysta2011}
\bibinfo{author}{\bibfnamefont{U.~M.} \bibnamefont{Ognysta}},
  \bibinfo{author}{\bibfnamefont{A.~B.} \bibnamefont{Nych}},
  \bibinfo{author}{\bibfnamefont{V.~A.} \bibnamefont{Uzunova}},
  \bibinfo{author}{\bibfnamefont{V.~M.} \bibnamefont{Pergamenschik}},
  \bibinfo{author}{\bibfnamefont{V.~G.} \bibnamefont{Nazarenko}},
  \bibinfo{author}{\bibfnamefont{M.}~\bibnamefont{\ifmmode~\check{S}\else
  \v{S}\fi{}karabot}}, \bibnamefont{and}
  \bibinfo{author}{\bibfnamefont{I.}~\bibnamefont{Mu\ifmmode \check{s}\else
  \v{s}\fi{}evi\ifmmode~\check{c}\else \v{c}\fi{}}}, \bibinfo{journal}{Phys.
  Rev. E} \textbf{\bibinfo{volume}{83}}, \bibinfo{pages}{041709}
  (\bibinfo{year}{2011}),
  \urlprefix\url{https://doi.org/10.1103/PhysRevE.83.041709}.

\bibitem[{\citenamefont{Tkalec and
  Mu{\v{s}}evi{\v{c}}}(2013)}]{Tkalec2013topology}
\bibinfo{author}{\bibfnamefont{U.}~\bibnamefont{Tkalec}} \bibnamefont{and}
  \bibinfo{author}{\bibfnamefont{I.}~\bibnamefont{Mu{\v{s}}evi{\v{c}}}},
  \bibinfo{journal}{Soft Matter} \textbf{\bibinfo{volume}{9}},
  \bibinfo{pages}{8140} (\bibinfo{year}{2013}),
  \urlprefix\url{https://doi.org/10.1039/c3sm50713k}.

\bibitem[{\citenamefont{Wang et~al.}(2017)\citenamefont{Wang, Zhang, and
  Chen}}]{wang2017}
\bibinfo{author}{\bibfnamefont{Y.}~\bibnamefont{Wang}},
  \bibinfo{author}{\bibfnamefont{P.}~\bibnamefont{Zhang}}, \bibnamefont{and}
  \bibinfo{author}{\bibfnamefont{J.~Z.~Y.} \bibnamefont{Chen}},
  \bibinfo{journal}{Phys. Rev. E} \textbf{\bibinfo{volume}{96}},
  \bibinfo{pages}{042702} (\bibinfo{year}{2017}),
  \urlprefix\url{https://doi.org/10.1103/PhysRevE.96.042702}.

\bibitem[{\citenamefont{Wang et~al.}(2018{\natexlab{a}})\citenamefont{Wang,
  Zhang, and Chen}}]{Wang2018}
\bibinfo{author}{\bibfnamefont{Y.}~\bibnamefont{Wang}},
  \bibinfo{author}{\bibfnamefont{P.}~\bibnamefont{Zhang}}, \bibnamefont{and}
  \bibinfo{author}{\bibfnamefont{J.~Z.~Y.} \bibnamefont{Chen}},
  \bibinfo{journal}{Soft Matter} \textbf{\bibinfo{volume}{14}},
  \bibinfo{pages}{6756} (\bibinfo{year}{2018}{\natexlab{a}}),
  \urlprefix\url{http://doi.org/10.1039/C8SM01057A}.

\bibitem[{\citenamefont{Lubensky and Prost}(1992)}]{Lubensky1992}
\bibinfo{author}{\bibfnamefont{T.~C.} \bibnamefont{Lubensky}} \bibnamefont{and}
  \bibinfo{author}{\bibfnamefont{J.}~\bibnamefont{Prost}}, \bibinfo{journal}{J.
  Phys. II (France)} \textbf{\bibinfo{volume}{2}}, \bibinfo{pages}{371}
  (\bibinfo{year}{1992}), \urlprefix\url{http://doi.org/10.1051/jp2:1992133}.

\bibitem[{\citenamefont{Nelson}(2002,~392pp)}]{Nelson2002a}
\bibinfo{author}{\bibfnamefont{D.~R.} \bibnamefont{Nelson}},
  \emph{\bibinfo{title}{{Defects and geometry in condensed matter physics}}}
  (\bibinfo{publisher}{Cambridge University Press},
  \bibinfo{year}{2002,~392pp}).

\bibitem[{\citenamefont{Arsenault et~al.}(2004)\citenamefont{Arsenault,
  Fournier-Bidoz, Hatton, Miguez, Tetreault, Vekris, Wong, Ming~Yang, Kitaev,
  and Ozin}}]{arsenault2004towards}
\bibinfo{author}{\bibfnamefont{A.}~\bibnamefont{Arsenault}},
  \bibinfo{author}{\bibfnamefont{S.}~\bibnamefont{Fournier-Bidoz}},
  \bibinfo{author}{\bibfnamefont{B.}~\bibnamefont{Hatton}},
  \bibinfo{author}{\bibfnamefont{H.}~\bibnamefont{Miguez}},
  \bibinfo{author}{\bibfnamefont{N.}~\bibnamefont{Tetreault}},
  \bibinfo{author}{\bibfnamefont{E.}~\bibnamefont{Vekris}},
  \bibinfo{author}{\bibfnamefont{S.}~\bibnamefont{Wong}},
  \bibinfo{author}{\bibfnamefont{S.}~\bibnamefont{Ming~Yang}},
  \bibinfo{author}{\bibfnamefont{V.}~\bibnamefont{Kitaev}}, \bibnamefont{and}
  \bibinfo{author}{\bibfnamefont{G.~A.} \bibnamefont{Ozin}},
  \bibinfo{journal}{J. Mater. Chem.} \textbf{\bibinfo{volume}{14}},
  \bibinfo{pages}{781} (\bibinfo{year}{2004}),
  \urlprefix\url{http://doi.org/10.1039/B314305H}.

\bibitem[{\citenamefont{Li et~al.}(2009)\citenamefont{Li, Yoo, Beernink, and
  Stein}}]{li2009site}
\bibinfo{author}{\bibfnamefont{F.}~\bibnamefont{Li}},
  \bibinfo{author}{\bibfnamefont{W.~C.} \bibnamefont{Yoo}},
  \bibinfo{author}{\bibfnamefont{M.~B.} \bibnamefont{Beernink}},
  \bibnamefont{and} \bibinfo{author}{\bibfnamefont{A.}~\bibnamefont{Stein}},
  \bibinfo{journal}{J. Am. Chem. Soc.} \textbf{\bibinfo{volume}{131}},
  \bibinfo{pages}{18548} (\bibinfo{year}{2009}),
  \urlprefix\url{http://doi.org/10.1021/ja908364k}.

\bibitem[{\citenamefont{Bowick and Giomi}(2009)}]{Bowick2009}
\bibinfo{author}{\bibfnamefont{M.~J.} \bibnamefont{Bowick}} \bibnamefont{and}
  \bibinfo{author}{\bibfnamefont{L.}~\bibnamefont{Giomi}},
  \bibinfo{journal}{Adv. Phys.} \textbf{\bibinfo{volume}{58}},
  \bibinfo{pages}{449} (\bibinfo{year}{2009}),
  \urlprefix\url{https://doi.org/10.1080/00018730903043166}.

\bibitem[{\citenamefont{Fern\'andez-Nieves
  et~al.}(2007)\citenamefont{Fern\'andez-Nieves, Vitelli, Utada, Link,
  M\'arquez, Nelson, and Weitz}}]{fernandez2007noval}
\bibinfo{author}{\bibfnamefont{A.}~\bibnamefont{Fern\'andez-Nieves}},
  \bibinfo{author}{\bibfnamefont{V.}~\bibnamefont{Vitelli}},
  \bibinfo{author}{\bibfnamefont{A.~S.} \bibnamefont{Utada}},
  \bibinfo{author}{\bibfnamefont{D.~R.} \bibnamefont{Link}},
  \bibinfo{author}{\bibfnamefont{M.}~\bibnamefont{M\'arquez}},
  \bibinfo{author}{\bibfnamefont{D.~R.} \bibnamefont{Nelson}},
  \bibnamefont{and} \bibinfo{author}{\bibfnamefont{D.~A.} \bibnamefont{Weitz}},
  \bibinfo{journal}{Phys. Rev. Lett.} \textbf{\bibinfo{volume}{99}},
  \bibinfo{pages}{157801} (\bibinfo{year}{2007}),
  \urlprefix\url{http://doi.org/10.1103/PhysRevLett.99.157801}.

\bibitem[{\citenamefont{Lopez-Leon et~al.}(2011)\citenamefont{Lopez-Leon,
  Koning, Devaiah, Vitelli, and Fernandez-Nieves}}]{Lopez-Leon2011}
\bibinfo{author}{\bibfnamefont{T.}~\bibnamefont{Lopez-Leon}},
  \bibinfo{author}{\bibfnamefont{V.}~\bibnamefont{Koning}},
  \bibinfo{author}{\bibfnamefont{K.~B.~S.} \bibnamefont{Devaiah}},
  \bibinfo{author}{\bibfnamefont{V.}~\bibnamefont{Vitelli}}, \bibnamefont{and}
  \bibinfo{author}{\bibfnamefont{A.}~\bibnamefont{Fernandez-Nieves}},
  \bibinfo{journal}{Nat. Phys.} \textbf{\bibinfo{volume}{7}},
  \bibinfo{pages}{391} (\bibinfo{year}{2011}),
  \urlprefix\url{http://doi.org/10.1038/nphys1920}.

\bibitem[{\citenamefont{Nelson}(2002)}]{nelson2002toward}
\bibinfo{author}{\bibfnamefont{D.~R.} \bibnamefont{Nelson}},
  \bibinfo{journal}{Nano Lett.} \textbf{\bibinfo{volume}{2}},
  \bibinfo{pages}{1125} (\bibinfo{year}{2002}),
  \urlprefix\url{http://doi.org/10.1021/nl0202096}.

\bibitem[{\citenamefont{Huber and Stark}(2005)}]{huber2005}
\bibinfo{author}{\bibfnamefont{M.}~\bibnamefont{Huber}} \bibnamefont{and}
  \bibinfo{author}{\bibfnamefont{H.}~\bibnamefont{Stark}},
  \bibinfo{journal}{EPL (Europhys. Lett.)} \textbf{\bibinfo{volume}{69}},
  \bibinfo{pages}{135} (\bibinfo{year}{2005}),
  \urlprefix\url{http://stacks.iop.org/0295-5075/69/i=1/a=135}.

\bibitem[{\citenamefont{Ska{\v c}ej and Zannoni}(2008)}]{skacej2008controlling}
\bibinfo{author}{\bibfnamefont{G.}~\bibnamefont{Ska{\v c}ej}} \bibnamefont{and}
  \bibinfo{author}{\bibfnamefont{C.}~\bibnamefont{Zannoni}},
  \bibinfo{journal}{Phys. Rev. Lett.} \textbf{\bibinfo{volume}{100}},
  \bibinfo{pages}{197802} (\bibinfo{year}{2008}),
  \urlprefix\url{http://link.aps.org/doi/10.1103/PhysRevLett.100.197802}.

\bibitem[{\citenamefont{Shin et~al.}(2008)\citenamefont{Shin, Bowick, and
  Xing}}]{Shin2008}
\bibinfo{author}{\bibfnamefont{H.}~\bibnamefont{Shin}},
  \bibinfo{author}{\bibfnamefont{M.}~\bibnamefont{Bowick}}, \bibnamefont{and}
  \bibinfo{author}{\bibfnamefont{X.}~\bibnamefont{Xing}},
  \bibinfo{journal}{Phys. Rev. Lett.} \textbf{\bibinfo{volume}{101}},
  \bibinfo{pages}{037802} (\bibinfo{year}{2008}),
  \urlprefix\url{http://doi.org/10.1103/PhysRevLett.101.037802}.

\bibitem[{\citenamefont{Bates}(2008)}]{Bates2008}
\bibinfo{author}{\bibfnamefont{M.~A.} \bibnamefont{Bates}},
  \bibinfo{journal}{J. Chem. Phys.} \textbf{\bibinfo{volume}{128}},
  \bibinfo{pages}{104707} (\bibinfo{year}{2008}),
  \urlprefix\url{http://doi.org/10.1063/1.2890724}.

\bibitem[{\citenamefont{Dhakal et~al.}(2012)\citenamefont{Dhakal, Solis, and
  Olvera de~la Cruz}}]{dhakal2002nematic}
\bibinfo{author}{\bibfnamefont{S.}~\bibnamefont{Dhakal}},
  \bibinfo{author}{\bibfnamefont{F.~J.} \bibnamefont{Solis}}, \bibnamefont{and}
  \bibinfo{author}{\bibfnamefont{M.}~\bibnamefont{Olvera de~la Cruz}},
  \bibinfo{journal}{Phys. Rev. E} \textbf{\bibinfo{volume}{86}},
  \bibinfo{pages}{011709} (\bibinfo{year}{2012}),
  \urlprefix\url{http://link.aps.org/doi/10.1103/PhysRevE.86.011709}.

\bibitem[{\citenamefont{Zhang et~al.}(2012{\natexlab{a}})\citenamefont{Zhang,
  Jiang, and Chen}}]{Zhang2012prl}
\bibinfo{author}{\bibfnamefont{W.-Y.} \bibnamefont{Zhang}},
  \bibinfo{author}{\bibfnamefont{Y.}~\bibnamefont{Jiang}}, \bibnamefont{and}
  \bibinfo{author}{\bibfnamefont{J.~Z.~Y.} \bibnamefont{Chen}},
  \bibinfo{journal}{Phys. Rev. Lett.} \textbf{\bibinfo{volume}{108}},
  \bibinfo{pages}{057801} (\bibinfo{year}{2012}{\natexlab{a}}),
  \urlprefix\url{http://link.aps.org/doi/10.1103/PhysRevLett.108.057801}.

\bibitem[{\citenamefont{Zhang et~al.}(2012{\natexlab{b}})\citenamefont{Zhang,
  Jiang, and Chen}}]{Zhang2012}
\bibinfo{author}{\bibfnamefont{W.-Y.} \bibnamefont{Zhang}},
  \bibinfo{author}{\bibfnamefont{Y.}~\bibnamefont{Jiang}}, \bibnamefont{and}
  \bibinfo{author}{\bibfnamefont{J.~Z.~Y.} \bibnamefont{Chen}},
  \bibinfo{journal}{Phys. Rev. E} \textbf{\bibinfo{volume}{85}},
  \bibinfo{pages}{061710} (\bibinfo{year}{2012}{\natexlab{b}}),
  \urlprefix\url{http://link.aps.org/doi/10.1103/PhysRevE.85.061710}.

\bibitem[{\citenamefont{Li et~al.}(2013)\citenamefont{Li, Miao, Ma, and
  Chen}}]{LiMiaoMaChen2013}
\bibinfo{author}{\bibfnamefont{Y.}~\bibnamefont{Li}},
  \bibinfo{author}{\bibfnamefont{H.}~\bibnamefont{Miao}},
  \bibinfo{author}{\bibfnamefont{H.}~\bibnamefont{Ma}}, \bibnamefont{and}
  \bibinfo{author}{\bibfnamefont{J.~Z.~Y.} \bibnamefont{Chen}},
  \bibinfo{journal}{Soft Matter} \textbf{\bibinfo{volume}{9}},
  \bibinfo{pages}{11461} (\bibinfo{year}{2013}),
  \urlprefix\url{http://doi.org/10.1039/c3sm52394b}.

\bibitem[{\citenamefont{Liang et~al.}(2014)\citenamefont{Liang, Ye, Zhang, and
  Chen}}]{Liang2014}
\bibinfo{author}{\bibfnamefont{Q.}~\bibnamefont{Liang}},
  \bibinfo{author}{\bibfnamefont{S.}~\bibnamefont{Ye}},
  \bibinfo{author}{\bibfnamefont{P.}~\bibnamefont{Zhang}}, \bibnamefont{and}
  \bibinfo{author}{\bibfnamefont{J.~Z.~Y.} \bibnamefont{Chen}},
  \bibinfo{journal}{J. Chem. Phys.} \textbf{\bibinfo{volume}{141}},
  \bibinfo{pages}{244901} (\bibinfo{year}{2014}),
  \urlprefix\url{http://doi.org/10.1063/1.4903995}.

\bibitem[{\citenamefont{Evans}(1995)}]{Evans1995}
\bibinfo{author}{\bibfnamefont{R.~M.} \bibnamefont{Evans}},
  \bibinfo{journal}{J. Phys. II} \textbf{\bibinfo{volume}{5}},
  \bibinfo{pages}{507} (\bibinfo{year}{1995}),
  \urlprefix\url{http://doi.org/10.1051/jp2:1995147}.

\bibitem[{\citenamefont{Bowick et~al.}(2004)\citenamefont{Bowick, Nelson, and
  Travesset}}]{Bowick2004}
\bibinfo{author}{\bibfnamefont{M.}~\bibnamefont{Bowick}},
  \bibinfo{author}{\bibfnamefont{D.~R.} \bibnamefont{Nelson}},
  \bibnamefont{and}
  \bibinfo{author}{\bibfnamefont{A.}~\bibnamefont{Travesset}},
  \bibinfo{journal}{Phys. Rev. E: Stat., Nonlinear, Soft Matter Phys.}
  \textbf{\bibinfo{volume}{69}}, \bibinfo{pages}{41102} (\bibinfo{year}{2004}),
  \urlprefix\url{http://doi.org/10.1103/PhysRevE.69.041102}.

\bibitem[{\citenamefont{Selinger et~al.}(2011)\citenamefont{Selinger, Konya,
  Travesset, and Selinger}}]{Selinger2011}
\bibinfo{author}{\bibfnamefont{R.~L.~B.} \bibnamefont{Selinger}},
  \bibinfo{author}{\bibfnamefont{A.}~\bibnamefont{Konya}},
  \bibinfo{author}{\bibfnamefont{A.}~\bibnamefont{Travesset}},
  \bibnamefont{and} \bibinfo{author}{\bibfnamefont{J.~V.}
  \bibnamefont{Selinger}}, \bibinfo{journal}{J. Phys. Chem. B}
  \textbf{\bibinfo{volume}{115}}, \bibinfo{pages}{13989}
  (\bibinfo{year}{2011}), \urlprefix\url{http://doi.org/10.1021/jp205128g}.

\bibitem[{\citenamefont{Li et~al.}(2014)\citenamefont{Li, Miao, Ma, and
  Chen}}]{LiYao2014}
\bibinfo{author}{\bibfnamefont{Y.}~\bibnamefont{Li}},
  \bibinfo{author}{\bibfnamefont{H.}~\bibnamefont{Miao}},
  \bibinfo{author}{\bibfnamefont{H.}~\bibnamefont{Ma}}, \bibnamefont{and}
  \bibinfo{author}{\bibfnamefont{J.~Z.~Y.} \bibnamefont{Chen}},
  \bibinfo{journal}{RSC Adv.} \textbf{\bibinfo{volume}{4}},
  \bibinfo{pages}{27471} (\bibinfo{year}{2014}),
  \urlprefix\url{http://doi.org/10.1039/C4RA04441J}.

\bibitem[{\citenamefont{Segatti et~al.}(2014)\citenamefont{Segatti, Snarski,
  and Veneroni}}]{Segatti2014}
\bibinfo{author}{\bibfnamefont{A.}~\bibnamefont{Segatti}},
  \bibinfo{author}{\bibfnamefont{M.}~\bibnamefont{Snarski}}, \bibnamefont{and}
  \bibinfo{author}{\bibfnamefont{M.}~\bibnamefont{Veneroni}},
  \bibinfo{journal}{Phys. Rev. E} \textbf{\bibinfo{volume}{90}},
  \bibinfo{pages}{012501} (\bibinfo{year}{2014}),
  \urlprefix\url{http://link.aps.org/doi/10.1103/PhysRevE.90.012501}.

\bibitem[{\citenamefont{Jesenek et~al.}(2015)\citenamefont{Jesenek, Kralj,
  Rosso, and Virga}}]{Jesenek2015}
\bibinfo{author}{\bibfnamefont{D.}~\bibnamefont{Jesenek}},
  \bibinfo{author}{\bibfnamefont{S.}~\bibnamefont{Kralj}},
  \bibinfo{author}{\bibfnamefont{R.}~\bibnamefont{Rosso}}, \bibnamefont{and}
  \bibinfo{author}{\bibfnamefont{E.~G.} \bibnamefont{Virga}},
  \bibinfo{journal}{Soft Matter} \textbf{\bibinfo{volume}{11}},
  \bibinfo{pages}{2434} (\bibinfo{year}{2015}),
  \urlprefix\url{http://doi.org/10.1039/C4SM02540G}.

\bibitem[{\citenamefont{Sheng}(1976)}]{Sheng1976}
\bibinfo{author}{\bibfnamefont{P.}~\bibnamefont{Sheng}},
  \bibinfo{journal}{Phys. Rev. Lett.} \textbf{\bibinfo{volume}{37}},
  \bibinfo{pages}{1059} (\bibinfo{year}{1976}),
  \urlprefix\url{http://link.aps.org/doi/10.1103/PhysRevLett.37.1059}.

\bibitem[{\citenamefont{Sheng}(1982)}]{Sheng1982}
\bibinfo{author}{\bibfnamefont{P.}~\bibnamefont{Sheng}},
  \bibinfo{journal}{Phys. Rev. A} \textbf{\bibinfo{volume}{26}},
  \bibinfo{pages}{1610} (\bibinfo{year}{1982}),
  \urlprefix\url{http://link.aps.org/doi/10.1103/PhysRevA.26.1610}.

\bibitem[{\citenamefont{Napoli and Vergori}(2012{\natexlab{a}})}]{Napoli2012}
\bibinfo{author}{\bibfnamefont{G.}~\bibnamefont{Napoli}} \bibnamefont{and}
  \bibinfo{author}{\bibfnamefont{L.}~\bibnamefont{Vergori}},
  \bibinfo{journal}{Phys. Rev. Lett.} \textbf{\bibinfo{volume}{108}},
  \bibinfo{pages}{207803} (\bibinfo{year}{2012}{\natexlab{a}}),
  \urlprefix\url{http://link.aps.org/doi/10.1103/PhysRevLett.108.207803}.

\bibitem[{\citenamefont{Napoli and Vergori}(2012{\natexlab{b}})}]{Napoli2012a}
\bibinfo{author}{\bibfnamefont{G.}~\bibnamefont{Napoli}} \bibnamefont{and}
  \bibinfo{author}{\bibfnamefont{L.}~\bibnamefont{Vergori}},
  \bibinfo{journal}{Phys. Rev. E} \textbf{\bibinfo{volume}{85}},
  \bibinfo{pages}{061701} (\bibinfo{year}{2012}{\natexlab{b}}),
  \urlprefix\url{http://link.aps.org/doi/10.1103/PhysRevE.85.061701}.

\bibitem[{\citenamefont{Chrzanowska et~al.}(2001)\citenamefont{Chrzanowska,
  Teixeira, Ehrentraut, and Cleaver}}]{Chrzanowska2001}
\bibinfo{author}{\bibfnamefont{A.}~\bibnamefont{Chrzanowska}},
  \bibinfo{author}{\bibfnamefont{P.~I.~C.} \bibnamefont{Teixeira}},
  \bibinfo{author}{\bibfnamefont{H.}~\bibnamefont{Ehrentraut}},
  \bibnamefont{and} \bibinfo{author}{\bibfnamefont{D.~J.}
  \bibnamefont{Cleaver}}, \bibinfo{journal}{J. Phys. Condens. Matter}
  \textbf{\bibinfo{volume}{13}}, \bibinfo{pages}{4715} (\bibinfo{year}{2001}),
  \urlprefix\url{http://doi.org/10.1088/0953-8984/13/21/306}.

\bibitem[{\citenamefont{Chrzanowska}(2003)}]{Chrzanowska2003}
\bibinfo{author}{\bibfnamefont{A.}~\bibnamefont{Chrzanowska}},
  \bibinfo{journal}{J. Comput. Phys.} \textbf{\bibinfo{volume}{191}},
  \bibinfo{pages}{265} (\bibinfo{year}{2003}),
  \urlprefix\url{http://doi.org/10.1016/S0021-9991(03)00316-4}.

\bibitem[{\citenamefont{de~las Heras et~al.}(2004)\citenamefont{de~las Heras,
  Velasco, and Mederos}}]{delasHeras2004}
\bibinfo{author}{\bibfnamefont{D.}~\bibnamefont{de~las Heras}},
  \bibinfo{author}{\bibfnamefont{E.}~\bibnamefont{Velasco}}, \bibnamefont{and}
  \bibinfo{author}{\bibfnamefont{L.}~\bibnamefont{Mederos}},
  \bibinfo{journal}{J. Chem. Phys.} \textbf{\bibinfo{volume}{120}},
  \bibinfo{pages}{4949} (\bibinfo{year}{2004}).

\bibitem[{\citenamefont{de~las Heras et~al.}(2009)\citenamefont{de~las Heras,
  Velasco, and Mederos}}]{delasHeras2009}
\bibinfo{author}{\bibfnamefont{D.}~\bibnamefont{de~las Heras}},
  \bibinfo{author}{\bibfnamefont{E.}~\bibnamefont{Velasco}}, \bibnamefont{and}
  \bibinfo{author}{\bibfnamefont{L.}~\bibnamefont{Mederos}},
  \bibinfo{journal}{Phys. Rev. E} \textbf{\bibinfo{volume}{79}},
  \bibinfo{pages}{061703} (\bibinfo{year}{2009}),
  \urlprefix\url{http://link.aps.org/doi/10.1103/PhysRevE.79.061703}.

\bibitem[{\citenamefont{Emelyanenko et~al.}(2011)\citenamefont{Emelyanenko,
  Aya, Sasaki, Araoka, Ema, Ishikawa, and Takezoe}}]{Emelyanenko2011}
\bibinfo{author}{\bibfnamefont{A.~V.} \bibnamefont{Emelyanenko}},
  \bibinfo{author}{\bibfnamefont{S.}~\bibnamefont{Aya}},
  \bibinfo{author}{\bibfnamefont{Y.}~\bibnamefont{Sasaki}},
  \bibinfo{author}{\bibfnamefont{F.}~\bibnamefont{Araoka}},
  \bibinfo{author}{\bibfnamefont{K.}~\bibnamefont{Ema}},
  \bibinfo{author}{\bibfnamefont{K.}~\bibnamefont{Ishikawa}}, \bibnamefont{and}
  \bibinfo{author}{\bibfnamefont{H.}~\bibnamefont{Takezoe}},
  \bibinfo{journal}{Phys. Rev. E} \textbf{\bibinfo{volume}{84}},
  \bibinfo{pages}{041701} (\bibinfo{year}{2011}),
  \urlprefix\url{http://link.aps.org/doi/10.1103/PhysRevE.84.041701}.

\bibitem[{\citenamefont{Chen}(2013)}]{Chen2013}
\bibinfo{author}{\bibfnamefont{J.~Z.~Y.} \bibnamefont{Chen}},
  \bibinfo{journal}{Soft Matter} \textbf{\bibinfo{volume}{9}},
  \bibinfo{pages}{10921} (\bibinfo{year}{2013}),
  \urlprefix\url{https://doi.org/10.1039/C3SM51991K}.

\bibitem[{\citenamefont{Tsakonas et~al.}(2007)\citenamefont{Tsakonas, Davidson,
  Brown, and Mottram}}]{Tsakonas2007}
\bibinfo{author}{\bibfnamefont{C.}~\bibnamefont{Tsakonas}},
  \bibinfo{author}{\bibfnamefont{A.~J.} \bibnamefont{Davidson}},
  \bibinfo{author}{\bibfnamefont{C.~V.} \bibnamefont{Brown}}, \bibnamefont{and}
  \bibinfo{author}{\bibfnamefont{N.~J.} \bibnamefont{Mottram}},
  \bibinfo{journal}{Appl. Phys. Lett.} \textbf{\bibinfo{volume}{90}},
  \bibinfo{pages}{111913} (\bibinfo{year}{2007}),
  \urlprefix\url{http://doi.org/10.1063/1.2713140}.

\bibitem[{\citenamefont{Galanis et~al.}(2006)\citenamefont{Galanis, Harries,
  Dan, Losert, and Nossal}}]{Galanis2006Spontaneous}
\bibinfo{author}{\bibfnamefont{J.}~\bibnamefont{Galanis}},
  \bibinfo{author}{\bibfnamefont{D.}~\bibnamefont{Harries}},
  \bibinfo{author}{\bibfnamefont{L.~S.} \bibnamefont{Dan}},
  \bibinfo{author}{\bibfnamefont{W.}~\bibnamefont{Losert}}, \bibnamefont{and}
  \bibinfo{author}{\bibfnamefont{R.}~\bibnamefont{Nossal}},
  \bibinfo{journal}{Phys. Rev. Lett.} \textbf{\bibinfo{volume}{96}},
  \bibinfo{pages}{028002} (\bibinfo{year}{2006}),
  \urlprefix\url{http://doi.org/10.1103/PhysRevLett.96.028002}.

\bibitem[{\citenamefont{Cortes et~al.}(2017)\citenamefont{Cortes, Gao, Dullens,
  and Aarts}}]{Cortes2017}
\bibinfo{author}{\bibfnamefont{L.~B.~G.} \bibnamefont{Cortes}},
  \bibinfo{author}{\bibfnamefont{Y.}~\bibnamefont{Gao}},
  \bibinfo{author}{\bibfnamefont{R.~P.~A.} \bibnamefont{Dullens}},
  \bibnamefont{and} \bibinfo{author}{\bibfnamefont{D.~G. A.~L.}
  \bibnamefont{Aarts}}, \bibinfo{journal}{J. Phys. Condens. Matter}
  \textbf{\bibinfo{volume}{29}}, \bibinfo{pages}{064003}
  (\bibinfo{year}{2017}),
  \urlprefix\url{http://doi.org/10.1088/1361-648X/29/6/064003}.

\bibitem[{\citenamefont{Wittmann et~al.}(2021)\citenamefont{Wittmann, Cortes,
  L{\"o}wen, and Aarts}}]{Aarts2021Particle}
\bibinfo{author}{\bibfnamefont{R.}~\bibnamefont{Wittmann}},
  \bibinfo{author}{\bibfnamefont{L.~B.~G.} \bibnamefont{Cortes}},
  \bibinfo{author}{\bibfnamefont{H.}~\bibnamefont{L{\"o}wen}},
  \bibnamefont{and} \bibinfo{author}{\bibfnamefont{D.~G. A.~L.}
  \bibnamefont{Aarts}}, \bibinfo{journal}{Nat. Commun.}
  \textbf{\bibinfo{volume}{12}}, \bibinfo{pages}{623} (\bibinfo{year}{2021}),
  \urlprefix\url{http://doi.org/10.1038/s41467-020-20842-5}.

\bibitem[{\citenamefont{Soares~e Silva et~al.}(2011)\citenamefont{Soares~e
  Silva, Alvarado, Nguyen, Georgoulia, Mulder, and Koenderink}}]{Mulder2011}
\bibinfo{author}{\bibfnamefont{M.}~\bibnamefont{Soares~e Silva}},
  \bibinfo{author}{\bibfnamefont{J.}~\bibnamefont{Alvarado}},
  \bibinfo{author}{\bibfnamefont{J.}~\bibnamefont{Nguyen}},
  \bibinfo{author}{\bibfnamefont{N.}~\bibnamefont{Georgoulia}},
  \bibinfo{author}{\bibfnamefont{B.~M.} \bibnamefont{Mulder}},
  \bibnamefont{and} \bibinfo{author}{\bibfnamefont{G.~H.}
  \bibnamefont{Koenderink}}, \bibinfo{journal}{Soft Matter}
  \textbf{\bibinfo{volume}{7}}, \bibinfo{pages}{10631} (\bibinfo{year}{2011}),
  \urlprefix\url{http://doi.org/10.1039/C1SM06060K}.

\bibitem[{\citenamefont{Lewis et~al.}(2014)\citenamefont{Lewis, Garlea,
  Alvarado, Dammone, Howell, Majumdar, Mulder, Lettinga, Koenderink, and
  Aarts}}]{Lewis2014}
\bibinfo{author}{\bibfnamefont{A.~H.} \bibnamefont{Lewis}},
  \bibinfo{author}{\bibfnamefont{I.}~\bibnamefont{Garlea}},
  \bibinfo{author}{\bibfnamefont{J.}~\bibnamefont{Alvarado}},
  \bibinfo{author}{\bibfnamefont{O.~J.} \bibnamefont{Dammone}},
  \bibinfo{author}{\bibfnamefont{P.~D.} \bibnamefont{Howell}},
  \bibinfo{author}{\bibfnamefont{A.}~\bibnamefont{Majumdar}},
  \bibinfo{author}{\bibfnamefont{B.~M.} \bibnamefont{Mulder}},
  \bibinfo{author}{\bibfnamefont{M.~P.} \bibnamefont{Lettinga}},
  \bibinfo{author}{\bibfnamefont{G.~H.} \bibnamefont{Koenderink}},
  \bibnamefont{and} \bibinfo{author}{\bibfnamefont{D.~G. A.~L.}
  \bibnamefont{Aarts}}, \bibinfo{journal}{Soft Matter}
  \textbf{\bibinfo{volume}{10}}, \bibinfo{pages}{7865} (\bibinfo{year}{2014}),
  \urlprefix\url{http://doi.org/10.1039/c4sm01123f}.

\bibitem[{\citenamefont{G{\^ a}rlea et~al.}(2019)\citenamefont{G{\^ a}rlea,
  Dammone, Alvarado, Notenboom, Jia, Koenderink, Aarts, Lettinga, and
  Mulder}}]{Garlea2019Colloidal}
\bibinfo{author}{\bibfnamefont{I.~C.} \bibnamefont{G{\^ a}rlea}},
  \bibinfo{author}{\bibfnamefont{O.}~\bibnamefont{Dammone}},
  \bibinfo{author}{\bibfnamefont{J.}~\bibnamefont{Alvarado}},
  \bibinfo{author}{\bibfnamefont{V.}~\bibnamefont{Notenboom}},
  \bibinfo{author}{\bibfnamefont{Y.}~\bibnamefont{Jia}},
  \bibinfo{author}{\bibfnamefont{G.~H.} \bibnamefont{Koenderink}},
  \bibinfo{author}{\bibfnamefont{D.~G. A.~L.} \bibnamefont{Aarts}},
  \bibinfo{author}{\bibfnamefont{M.~P.} \bibnamefont{Lettinga}},
  \bibnamefont{and} \bibinfo{author}{\bibfnamefont{B.~M.}
  \bibnamefont{Mulder}}, \bibinfo{journal}{Sci. Rep.}
  \textbf{\bibinfo{volume}{9}}, \bibinfo{pages}{20391} (\bibinfo{year}{2019}),
  \urlprefix\url{http://doi.org/10.1038/s41598-019-56729-9}.

\bibitem[{\citenamefont{Han et~al.}(2021{\natexlab{a}})\citenamefont{Han, Yin,
  Hu, Majumdar, and Zhang}}]{han2021}
\bibinfo{author}{\bibfnamefont{Y.}~\bibnamefont{Han}},
  \bibinfo{author}{\bibfnamefont{J.}~\bibnamefont{Yin}},
  \bibinfo{author}{\bibfnamefont{Y.}~\bibnamefont{Hu}},
  \bibinfo{author}{\bibfnamefont{A.}~\bibnamefont{Majumdar}}, \bibnamefont{and}
  \bibinfo{author}{\bibfnamefont{L.}~\bibnamefont{Zhang}},
  \bibinfo{journal}{preprint arXiv:2106.12521}
  (\bibinfo{year}{2021}{\natexlab{a}}).

\bibitem[{\citenamefont{Everts et~al.}(2016)\citenamefont{Everts, Punter,
  Samin, van~der Schoot, and van Roij}}]{Everts2016A}
\bibinfo{author}{\bibfnamefont{J.~C.} \bibnamefont{Everts}},
  \bibinfo{author}{\bibfnamefont{M.~T. J. J.~M.} \bibnamefont{Punter}},
  \bibinfo{author}{\bibfnamefont{S.}~\bibnamefont{Samin}},
  \bibinfo{author}{\bibfnamefont{P.}~\bibnamefont{van~der Schoot}},
  \bibnamefont{and} \bibinfo{author}{\bibfnamefont{R.}~\bibnamefont{van Roij}},
  \bibinfo{journal}{J. Chem. Phys.} \textbf{\bibinfo{volume}{144}},
  \bibinfo{pages}{194901} (\bibinfo{year}{2016}),
  \urlprefix\url{http://doi.org/10.1063/1.4948785}.

\bibitem[{\citenamefont{Robinson et~al.}(2017)\citenamefont{Robinson, Luo,
  Farrell, Erban, and Majumdar}}]{Majumdar2017}
\bibinfo{author}{\bibfnamefont{M.}~\bibnamefont{Robinson}},
  \bibinfo{author}{\bibfnamefont{C.}~\bibnamefont{Luo}},
  \bibinfo{author}{\bibfnamefont{P.~E.} \bibnamefont{Farrell}},
  \bibinfo{author}{\bibfnamefont{R.}~\bibnamefont{Erban}}, \bibnamefont{and}
  \bibinfo{author}{\bibfnamefont{A.}~\bibnamefont{Majumdar}},
  \bibinfo{journal}{Liq. Cryst.} \textbf{\bibinfo{volume}{14-15}},
  \bibinfo{pages}{2267} (\bibinfo{year}{2017}),
  \urlprefix\url{https://doi.org/10.1080/02678292.2017.1290284}.

\bibitem[{\citenamefont{Wang et~al.}(2018{\natexlab{b}})\citenamefont{Wang,
  Canevari, and Majumdar}}]{Wang2018Order}
\bibinfo{author}{\bibfnamefont{Y.}~\bibnamefont{Wang}},
  \bibinfo{author}{\bibfnamefont{G.}~\bibnamefont{Canevari}}, \bibnamefont{and}
  \bibinfo{author}{\bibfnamefont{A.}~\bibnamefont{Majumdar}},
  \bibinfo{journal}{SIAM J. Appl. Math.} \textbf{\bibinfo{volume}{79}},
  \bibinfo{pages}{1314} (\bibinfo{year}{2018}{\natexlab{b}}),
  \urlprefix\url{http://doi.org/10.1137/17M1179820}.

\bibitem[{\citenamefont{Yin et~al.}(2020)\citenamefont{Yin, Wang, Chen, Zhang,
  and Zhang}}]{Yin2020Construction}
\bibinfo{author}{\bibfnamefont{J.}~\bibnamefont{Yin}},
  \bibinfo{author}{\bibfnamefont{Y.}~\bibnamefont{Wang}},
  \bibinfo{author}{\bibfnamefont{J.~Z.~Y.} \bibnamefont{Chen}},
  \bibinfo{author}{\bibfnamefont{P.}~\bibnamefont{Zhang}}, \bibnamefont{and}
  \bibinfo{author}{\bibfnamefont{L.}~\bibnamefont{Zhang}},
  \bibinfo{journal}{Phys. Rev. Lett.} \textbf{\bibinfo{volume}{124}},
  \bibinfo{pages}{090601} (\bibinfo{year}{2020}),
  \urlprefix\url{http://doi.org/10.1103/PhysRevLett.124.090601}.

\bibitem[{\citenamefont{Han et~al.}(2020)\citenamefont{Han, Majumdar, and
  Zhang}}]{Han2020A}
\bibinfo{author}{\bibfnamefont{Y.}~\bibnamefont{Han}},
  \bibinfo{author}{\bibfnamefont{A.}~\bibnamefont{Majumdar}}, \bibnamefont{and}
  \bibinfo{author}{\bibfnamefont{L.}~\bibnamefont{Zhang}},
  \bibinfo{journal}{SIAM J. Appl. Math.} \textbf{\bibinfo{volume}{80}},
  \bibinfo{pages}{1678} (\bibinfo{year}{2020}),
  \urlprefix\url{http://doi.org/10.1137/19M1293156}.

\bibitem[{\citenamefont{Han et~al.}(2021{\natexlab{b}})\citenamefont{Han, Yin,
  Zhang, Majumdar, and Zhang}}]{Han2021Solution}
\bibinfo{author}{\bibfnamefont{Y.}~\bibnamefont{Han}},
  \bibinfo{author}{\bibfnamefont{J.}~\bibnamefont{Yin}},
  \bibinfo{author}{\bibfnamefont{P.}~\bibnamefont{Zhang}},
  \bibinfo{author}{\bibfnamefont{A.}~\bibnamefont{Majumdar}}, \bibnamefont{and}
  \bibinfo{author}{\bibfnamefont{L.}~\bibnamefont{Zhang}},
  \bibinfo{journal}{Nonlinearity} \textbf{\bibinfo{volume}{34}},
  \bibinfo{pages}{2048} (\bibinfo{year}{2021}{\natexlab{b}}),
  \urlprefix\url{http://doi.org/10.1088/1361-6544/abc5d4}.

\bibitem[{\citenamefont{Yao et~al.}(2018)\citenamefont{Yao, Zhang, and
  Chen}}]{Yao2018}
\bibinfo{author}{\bibfnamefont{X.}~\bibnamefont{Yao}},
  \bibinfo{author}{\bibfnamefont{H.}~\bibnamefont{Zhang}}, \bibnamefont{and}
  \bibinfo{author}{\bibfnamefont{J.~Z.~Y.} \bibnamefont{Chen}},
  \bibinfo{journal}{Phys. Rev. E} \textbf{\bibinfo{volume}{97}},
  \bibinfo{pages}{052707} (\bibinfo{year}{2018}),
  \urlprefix\url{https://link.aps.org/doi/10.1103/PhysRevE.97.052707}.

\bibitem[{\citenamefont{Yao and Chen}(2020)}]{Yao2020}
\bibinfo{author}{\bibfnamefont{X.}~\bibnamefont{Yao}} \bibnamefont{and}
  \bibinfo{author}{\bibfnamefont{J.~Z.~Y.} \bibnamefont{Chen}},
  \bibinfo{journal}{Phys. Rev. E} \textbf{\bibinfo{volume}{101}},
  \bibinfo{pages}{062706} (\bibinfo{year}{2020}),
  \urlprefix\url{https://link.aps.org/doi/10.1103/PhysRevE.101.062706}.

\bibitem[{\citenamefont{Dzubiella et~al.}(2000)\citenamefont{Dzubiella,
  Schmidt, and L{\"{o}}wen}}]{Dzubiella2000Topological}
\bibinfo{author}{\bibfnamefont{J.}~\bibnamefont{Dzubiella}},
  \bibinfo{author}{\bibfnamefont{M.}~\bibnamefont{Schmidt}}, \bibnamefont{and}
  \bibinfo{author}{\bibfnamefont{H.}~\bibnamefont{L{\"{o}}wen}},
  \bibinfo{journal}{Phys. Rev. E} \textbf{\bibinfo{volume}{62}},
  \bibinfo{pages}{5081} (\bibinfo{year}{2000}),
  \urlprefix\url{http://doi.org/10.1103/PhysRevE.62.5081}.

\bibitem[{\citenamefont{de~las Heras and Velasco}(2014)}]{Heras2014Domain}
\bibinfo{author}{\bibfnamefont{D.}~\bibnamefont{de~las Heras}}
  \bibnamefont{and} \bibinfo{author}{\bibfnamefont{E.}~\bibnamefont{Velasco}},
  \bibinfo{journal}{Soft Matter} \textbf{\bibinfo{volume}{10}},
  \bibinfo{pages}{1758} (\bibinfo{year}{2014}),
  \urlprefix\url{https://doi.org/10.1039/c3sm52650j}.

\bibitem[{\citenamefont{G{\^{a}}rlea and Mulder}(2015)}]{Mulder2015}
\bibinfo{author}{\bibfnamefont{I.~C.} \bibnamefont{G{\^{a}}rlea}}
  \bibnamefont{and} \bibinfo{author}{\bibfnamefont{B.~M.}
  \bibnamefont{Mulder}}, \bibinfo{journal}{Soft Matter}
  \textbf{\bibinfo{volume}{11}}, \bibinfo{pages}{608} (\bibinfo{year}{2015}),
  \urlprefix\url{http://doi.org/10.1039/C4SM02087A}.

\bibitem[{\citenamefont{G{\^{a}}rlea et~al.}(2016)\citenamefont{G{\^{a}}rlea,
  Mulder, Alvarado, Dammone, Aarts, Lettinga, Koenderink, and
  Mulder}}]{Garlea2016Finite}
\bibinfo{author}{\bibfnamefont{I.~C.} \bibnamefont{G{\^{a}}rlea}},
  \bibinfo{author}{\bibfnamefont{P.}~\bibnamefont{Mulder}},
  \bibinfo{author}{\bibfnamefont{J.}~\bibnamefont{Alvarado}},
  \bibinfo{author}{\bibfnamefont{O.}~\bibnamefont{Dammone}},
  \bibinfo{author}{\bibfnamefont{D.~G. A.~L.} \bibnamefont{Aarts}},
  \bibinfo{author}{\bibfnamefont{M.~P.} \bibnamefont{Lettinga}},
  \bibinfo{author}{\bibfnamefont{G.~H.} \bibnamefont{Koenderink}},
  \bibnamefont{and} \bibinfo{author}{\bibfnamefont{B.~M.}
  \bibnamefont{Mulder}}, \bibinfo{journal}{Nat. Commun.}
  \textbf{\bibinfo{volume}{7}}, \bibinfo{pages}{12112} (\bibinfo{year}{2016}),
  \urlprefix\url{http://doi.org/10.1038/ncomms12112}.

\bibitem[{\citenamefont{Hashemi}(2019{\natexlab{a}})}]{Hashemi2019Structure}
\bibinfo{author}{\bibfnamefont{S.}~\bibnamefont{Hashemi}},
  \bibinfo{journal}{Braz. J. Phys.} \textbf{\bibinfo{volume}{49}},
  \bibinfo{pages}{321} (\bibinfo{year}{2019}{\natexlab{a}}),
  \urlprefix\url{http://doi.org/10.1007/s13538-019-00657-6}.

\bibitem[{\citenamefont{Hashemi}(2019{\natexlab{b}})}]{Hashemi2019Finding}
\bibinfo{author}{\bibfnamefont{S.}~\bibnamefont{Hashemi}},
  \bibinfo{journal}{Braz. J. Phys.} \textbf{\bibinfo{volume}{49}},
  \bibinfo{pages}{44} (\bibinfo{year}{2019}{\natexlab{b}}),
  \urlprefix\url{https://doi.org/10.1007/s13538-018-0612-6}.

\bibitem[{\citenamefont{Onsager}(1949)}]{Onsager1949}
\bibinfo{author}{\bibfnamefont{L.}~\bibnamefont{Onsager}},
  \bibinfo{journal}{Ann. N. Y. Acad. Sci.} \textbf{\bibinfo{volume}{51}},
  \bibinfo{pages}{627} (\bibinfo{year}{1949}),
  \urlprefix\url{http://doi.org/10.1111/j.1749-6632.1949.tb27296.x}.

\bibitem[{\citenamefont{Galanis et~al.}(2010)\citenamefont{Galanis, Nossal,
  Losert, and Harries}}]{Galanis2010}
\bibinfo{author}{\bibfnamefont{J.}~\bibnamefont{Galanis}},
  \bibinfo{author}{\bibfnamefont{R.}~\bibnamefont{Nossal}},
  \bibinfo{author}{\bibfnamefont{W.}~\bibnamefont{Losert}}, \bibnamefont{and}
  \bibinfo{author}{\bibfnamefont{D.}~\bibnamefont{Harries}},
  \bibinfo{journal}{Phys. Rev. Lett.} \textbf{\bibinfo{volume}{105}},
  \bibinfo{pages}{168001} (\bibinfo{year}{2010}),
  \urlprefix\url{http://doi.org/10.1103/PhysRevLett.105.168001}.

\bibitem[{\citenamefont{Luo et~al.}(2012)\citenamefont{Luo, Majumdar, and
  Erban}}]{Luo2012}
\bibinfo{author}{\bibfnamefont{C.}~\bibnamefont{Luo}},
  \bibinfo{author}{\bibfnamefont{A.}~\bibnamefont{Majumdar}}, \bibnamefont{and}
  \bibinfo{author}{\bibfnamefont{R.}~\bibnamefont{Erban}},
  \bibinfo{journal}{Phys. Rev. E} \textbf{\bibinfo{volume}{85}},
  \bibinfo{pages}{061702} (\bibinfo{year}{2012}),
  \urlprefix\url{http://doi.org/10.1103/PhysRevE.85.061702}.

\bibitem[{\citenamefont{Geigenfeind et~al.}(2015)\citenamefont{Geigenfeind,
  Rosenzweig, Schmidt, and de~las Heras}}]{Geigenfein2015}
\bibinfo{author}{\bibfnamefont{T.}~\bibnamefont{Geigenfeind}},
  \bibinfo{author}{\bibfnamefont{S.}~\bibnamefont{Rosenzweig}},
  \bibinfo{author}{\bibfnamefont{M.}~\bibnamefont{Schmidt}}, \bibnamefont{and}
  \bibinfo{author}{\bibfnamefont{D.}~\bibnamefont{de~las Heras}},
  \bibinfo{journal}{J. Chem. Phys.} \textbf{\bibinfo{volume}{142}},
  \bibinfo{pages}{174701} (\bibinfo{year}{2015}),
  \urlprefix\url{http://doi.org/10.1063/1.4919307}.

\bibitem[{\citenamefont{Ericksen}(1991)}]{ericksen1991liquid}
\bibinfo{author}{\bibfnamefont{J.~L.} \bibnamefont{Ericksen}},
  \bibinfo{journal}{Ratio. Mech. An.} \textbf{\bibinfo{volume}{113}},
  \bibinfo{pages}{97} (\bibinfo{year}{1991}),
  \urlprefix\url{http://doi.org/10.4310/cag.2018.v26.n2.a5}.

\bibitem[{\citenamefont{Pathria}(1996)}]{pathriastatistical}
\bibinfo{author}{\bibfnamefont{R.~K.} \bibnamefont{Pathria}},
  \emph{\bibinfo{title}{Statistical Mechanics}}
  (\bibinfo{publisher}{Butterworth Heinemann}, \bibinfo{address}{Oxford},
  \bibinfo{year}{1996}).

\bibitem[{\citenamefont{Kayser and Raveche}(1978)}]{Kayser1978}
\bibinfo{author}{\bibfnamefont{R.~F.} \bibnamefont{Kayser}} \bibnamefont{and}
  \bibinfo{author}{\bibfnamefont{H.~J.} \bibnamefont{Raveche}},
  \bibinfo{journal}{Phys. Rev. A} \textbf{\bibinfo{volume}{17}},
  \bibinfo{pages}{2067} (\bibinfo{year}{1978}),
  \urlprefix\url{http://doi.org/10.1103/PhysRevA.17.2067}.

\bibitem[{\citenamefont{Cuesta et~al.}(1989)\citenamefont{Cuesta, Tejero, and
  Baus}}]{Cuesta1989}
\bibinfo{author}{\bibfnamefont{J.~A.} \bibnamefont{Cuesta}},
  \bibinfo{author}{\bibfnamefont{C.~F.} \bibnamefont{Tejero}},
  \bibnamefont{and} \bibinfo{author}{\bibfnamefont{M.}~\bibnamefont{Baus}},
  \bibinfo{journal}{Phys. Rev. A} \textbf{\bibinfo{volume}{39}},
  \bibinfo{pages}{6498} (\bibinfo{year}{1989}),
  \urlprefix\url{http://doi.org/10.1103/PhysRevA.39.6498}.

\bibitem[{\citenamefont{Chen}(1993)}]{Chen1993prl}
\bibinfo{author}{\bibfnamefont{Z.~Y.} \bibnamefont{Chen}},
  \bibinfo{journal}{Phys. Rev. Lett.} \textbf{\bibinfo{volume}{71}},
  \bibinfo{pages}{93} (\bibinfo{year}{1993}),
  \urlprefix\url{https://doi.org/10.1103/PhysRevLett.71.93}.

\bibitem[{\citenamefont{Chen}(2016)}]{Chen2016}
\bibinfo{author}{\bibfnamefont{J.~Z.~Y.} \bibnamefont{Chen}},
  \bibinfo{journal}{Prog. Polym. Sci.} \textbf{\bibinfo{volume}{54-55}},
  \bibinfo{pages}{3} (\bibinfo{year}{2016}),
  \urlprefix\url{http://doi.org/10.1016/j.progpolymsci.2015.09.002}.

\end{thebibliography}
\bibliographystyle{apsrev}




\end{document}